\documentclass[aps,twocolumn,prd,amsmath,footinbib,preprintnumbers,superscriptaddress]{revtex4}
\usepackage{graphicx}
\usepackage{dcolumn}
\usepackage{bm}
\usepackage{xcolor}
\usepackage{amsmath}
\usepackage{booktabs,multirow,tabularx}
\usepackage{subfig}
\usepackage[colorlinks]{hyperref}
\hypersetup{colorlinks=true, linkcolor=blue, filecolor=blue, urlcolor=blue, citecolor=blue}
\def\msbar{\overline{\mathrm{MS}}}
\def\sims{\mathrm{SIM}}

\begin{document}

\title{Soft Contributions Stabilize NNLO QCD Corrections to Quarkonium Production and Decay}

\author{Luca Maxia}
\author{Hua-Sheng Shao}
\author{Lukas Simon}
\author{Guoxing Wang}%
\affiliation{%
 Laboratoire de Physique Th\'eorique et Hautes Energies (LPTHE), UMR 7589, Sorbonne Universit\'e et CNRS, 4 place Jussieu, 75252 Paris Cedex 05, France
}%

\date{\today}

\begin{abstract}
Next-to-next-to-leading order (NNLO) QCD corrections to quarkonium production and decay are known to exhibit perturbative instabilities within non-relativistic QCD. We identify the origin of this problem and propose a simple remedy. Applying our approach to $S$-wave color-singlet quarkonium processes, we achieve substantially improved perturbative convergence and agreement with experimental data.
\end{abstract}

\maketitle

As the simplest class of hadrons consisting of a heavy quark-antiquark pair, quarkonium plays a special role in deepening our understanding of strong interactions in nuclear and particle physics. In contrast to light or open heavy-flavor hadrons, quarkonium production and decay processes are widely regarded as tractable within non-relativistic quantum field theory frameworks. For instance, non-relativistic QCD (NRQCD)~\cite{Bodwin:1994jh} is a rigorous low-energy effective field theory of QCD, obtained by integrating out degrees of freedom associated with energy scales higher than the heavy-quark mass $m_Q$. It therefore enables systematic improvements in the calculation of quarkonium production and decay rates and is, to date, the most widely adopted theoretical framework for quarkonium studies. 

While many open issues remain~\cite{Brambilla:2010cs,Lansberg:2019adr,Chapon:2020heu}, a prominent purely theoretical challenge has emerged over the past decade in NRQCD studies of quarkonium: the convergence of the perturbative expansion in the strong coupling $\alpha_s$ is increasingly called into question for a wide range of quarkonium processes. Thanks to advances in multi-loop techniques~\cite{Weinzierl:2022eaz,Blumlein:2022qci,Armadillo:2025mvu} over the last ten years, a significant number of quarkonium production and decay processes have been computed at next-to-next-to-leading order (NNLO) accuracy in $\alpha_s$. However, several common issues have been observed in theoretical predictions once NNLO QCD corrections are included. In particular, NNLO corrections are often large and negative, which can lead to negative and thus unphysical partial decay widths or cross sections. Typical examples include $\eta_c \to \gamma\gamma$~\cite{Czarnecki:2001zc,Feng:2015uha,Abreu:2022cco}, $J/\psi \to e^+e^-$~\cite{Czarnecki:1997vz,Beneke:1997jm,Czarnecki:2001zc,Beneke:2014qea,Feng:2022vvk}, $J/\psi \to \gamma\gamma\gamma$~\cite{Zeng:2026ois}, and $e^+e^- \to J/\psi J/\psi$~\cite{Sang:2023liy,Huang:2023pmn}. For other observables that remain positive, such as the $\gamma\gamma^* \to \eta_c$ transition form factor~\cite{Feng:2015uha,Babiarz:2025agk}, the inclusion of NNLO corrections significantly increases the renormalization-scale uncertainty compared to next-to-leading order (NLO), contrary to the usual expectation of improved perturbative stability. Moreover, due to the large size of NNLO corrections, noticeable differences can also arise in exclusive processes depending on whether the $\alpha_s$ expansion is truncated at the amplitude level or at the squared-amplitude level (see, e.g., refs.~\cite{Huang:2023pmn,Zeng:2026ois,Feng:2015uha}).

There may be several possible explanations for these issues, depending on the specific problem, such as a large value of $\alpha_s$, the presence of large logarithms, or the emergence of new enhanced channels~\cite{Rubin:2010xp,Shao:2018adj,Pagani:2020rsg}, each of which requires different mitigation strategies. In this letter, however, we argue that the conventional modified minimal subtraction ($\msbar$) scheme that has been used in the literature to calculate quarkonium rates in NRQCD at NNLO accuracy in $\alpha_s$ is not a good choice.

To better understand the problem, we first examine the partial decay width of $\eta_c \to \gamma\gamma$, which is a single-scale process and therefore free of logarithms arising from hierarchical scales. Its $\alpha_s$ expansion up to NNLO~\cite{Czarnecki:2001zc,Feng:2015uha,Abreu:2022cco} is
\begin{equation}
\frac{\Gamma_{\mathrm{NNLO}}(\eta_c\to\gamma\gamma)}{\Gamma_{\mathrm{LO}}(\eta_c\to\gamma\gamma)}=1-1.07\alpha_s-11.16\alpha_s^2\,,\label{eq:etac2aaNNLOnoSoft}
\end{equation}
where $\Gamma_{\mathrm{LO}}$ is the leading order (LO) width. We have used $n_f=3$ quark flavors and set the renormalization and NRQCD factorization scales to the charm-quark mass, $\mu_R = \mu_{\Lambda}= m_c$, in the above equation. The negative NNLO width can be attributed to the large and negative coefficient of $\alpha_s^2$, whose magnitude is more than one order larger than that of the $\alpha_s$ term. This issue appears generically in many quarkonium processes known at NNLO in the literature that we have examined (see the Supplemental Material~\ref{sec:SM}).

Such an issue has been broadly acknowledged in the literature, yet no fully satisfactory solution has been proposed so far. Nevertheless, there are several suggestive hints. In NNLO QED corrections to para-positronium decays into two photons~\cite{Czarnecki:1999ci,Abreu:2022vei}, the large and negative $\alpha^2$ coefficient in the short-distance coefficient (SDC) cancels against a large and positive contribution arising from $\mathcal{O}(\alpha^2)$ soft corrections to the wavefunction at the origin, yielding a final $\mathcal{O}(1)$ coefficient at NNLO. In addition, ref.~\cite{Chung:2020zqc} presents a method to compute wavefunctions at the origin within potential NRQCD (pNRQCD)~\cite{Pineda:1997bj, Brambilla:1999xf, Beneke:1999zr, Beneke:1999qg}. In this framework, large cancellations are also observed between the SDCs of $\eta_c\to\gamma\gamma$ and $J/\psi\to e^+e^-$ and the non-Coulombic contributions to the wavefunctions at the origin, which include both perturbative and non-perturbative (NP) effects. These cancellations are interpreted as a consequence of the elimination of renormalon ambiguities~\cite{Beneke:1998ui} in the NP QCD potential. Finally, ref.~\cite{Kiyo:2010jm} adopts a weak-coupling approach to compute quarkonium long-distance matrix elements (LDMEs) perturbatively, incorporating both Coulombic and non-Coulombic corrections. This framework has been used to predict the leptonic decay rate of the $\Upsilon(1S)$ meson up to three loops in QCD~\cite{Beneke:2014qea} and has traditionally been used to describe top-quark pair production near threshold (see, e.g., ref.~\cite{Hoang:2000yr} and references therein). However, the authors of ref.~\cite{Kiyo:2010jm} express some reservations regarding the applicability of the weak-coupling approach to the charmonium system.

Inspired by the para-positronium decay into two photons~\cite{Czarnecki:1999ci,Abreu:2022vei}, we propose a simple solution to address the aforementioned perturbative instability problem for $S$-wave color-singlet quarkonium states, restricting ourselves to leading power (LP) in the heavy-quark relative velocity $v$ expansion within NRQCD. We remind the reader that, within the NRQCD factorization framework, only the combination of SDCs and LDMEs is physical. At NNLO in QCD, the SDCs alone are infrared (IR) divergent; these divergences must be canceled by the renormalization of the LDMEs, in close analogy with the renormalization of parton distribution functions in perturbative QCD factorization. A common practice in the literature is to renormalize the LDMEs using the $\msbar$ scheme, which retains only the poles in dimensional regularization. This procedure introduces a dependence on the NRQCD factorization scale $\mu_{\Lambda}$ in the SDCs, which should be (partially) compensated by the renormalization group (RG) evolution of the LDMEs, leaving a residual higher-order dependence on $\mu_{\Lambda}$. Physically, $\mu_{\Lambda}$ represents the ultraviolet (UV) cutoff of NRQCD, whose natural value is of the order of the heavy-quark mass $m_Q$. However, we argue that the widely used $\msbar$ scheme is not optimal for studying quarkonium processes perturbatively within NRQCD. Using the $\msbar$ scheme, in which the finite term vanishes, is analogous to setting the finite soft contribution to the wavefunction at the origin to zero in the para-positronium case; in this case, the NNLO cancellation observed in refs.~\cite{Czarnecki:1999ci, Abreu:2022vei} would no longer be manifest.

Unlike positronium, quarkonium involves both perturbative and NP aspects that require distinct treatments. Even within the purely perturbative sector, Coulomb resummation is necessary to improve perturbative stability. However, as in any resummation procedure, ambiguities arise in deciding which contributions should be expanded to fixed order in $\alpha_s$ and which should be resummed to all orders in $\alpha_s$~\cite{AH:2022elh}. A proper determination of the fixed-order perturbative part in the renormalized LDMEs is therefore crucial for stabilizing quarkonium NNLO QCD corrections.

In dimensional regularization, it is understood that Coulomb singularities do not generate $1/\epsilon$ poles, where $\epsilon$ is the dimensional regulator with spacetime dimension $d = 4 - 2\epsilon$. In contrast, the $1/\epsilon$ divergences arising in the renormalization of LDMEs, as well as their associated explicit $\mu_{\Lambda}$ dependence, originate from soft contributions. For $S$-wave color-singlet quarkonium states, for which the LDMEs can be related to the squared wavefunctions at the origin, the spin-independent three-loop Coulomb corrections to the wavefunction are known~\cite{Melnikov:1998ug,Beneke:1999qg,Beneke:2005hg}. In addition, for both spin-singlet and spin-triplet states, spin-dependent NNLO soft (non-Coulombic) corrections have been computed within the pNRQCD framework~\cite{Melnikov:1998ug,Beneke:1999qg,Kniehl:2002yv,Penin:2004ay,Beneke:2005hg,Kiyo:2010jm}.

An important observation we make here is that only the short-distance (SD) soft contributions to the wavefunctions at the origin should be included up to NNLO in $\alpha_s$, while the long-distance (LD) contributions, including soft, Coulombic, and NP contributions, should remain resummed. In other words, the LD parts of the LDMEs are to be determined by solving the Schr\"odinger equation with a phenomenological or lattice-inspired potential, or by extracting them from experimental data, as is commonly done in quarkonium studies. The full non-Coulombic perturbative soft contribution to the renormalized wavefunction at the origin at $\mathcal{O}(\alpha_s^2)$ is given by~\cite{Kniehl:2002yv,Penin:2004ay,Beneke:2005hg,Kiyo:2010jm}
\begin{eqnarray}
    \delta\psi^{\mathrm{Soft}}_{s,n}(0)&=&\psi_{n}^{\mathrm{LO}}(0)\Bigg\{\dfrac{\alpha_s^2(\mu_S)}{2}\bigg[C_F^2\left(\dfrac{L_{n}}{2s+1}-\dfrac{15}{8n^2}\right.\nonumber\\
    &&\left.+\dfrac{4}{(2s+1)n}+\dfrac{5}{3}+\dfrac{7}{3(2s+1)}-\dfrac{2S_1(n)}{2s+1}\right)\nonumber\\
    &&+C_AC_F\left(\dfrac{L_{n}}{2}+\dfrac{2}{n}+\dfrac{5}{4}-S_1(n)\right)\bigg]\Bigg\}\label{eq:softexpandforWFatoriginCS}
\end{eqnarray}
for an $S$-wave quarkonium state with spin $s \in \left\{0,1\right\}$. Here, $C_F=4/3$ and $C_A=3$ are the QCD Casimir factors, $n$ is the principal quantum number, $\psi_{n}^{\mathrm{LO}}(0)$ denotes the LO wavefunction at the origin, and $S_a(n)=\sum_{k=1}^{n} k^{-a}$ is the harmonic sum. The logarithm $L_n$ is defined as $L_n\equiv \log{\left(\mu_{\Lambda}^2a_n^2\right)}$, with the Bohr radius $a_n=n/(C_F\alpha_s(\mu_S)m_Q)$. The soft scale $\mu_S$ should be of order the inverse Bohr radius. At NNLO, the non-Coulombic correction to the wavefunction at the origin, $\delta\psi_{s,n}^{\mathrm{Soft}}(0)$, is generated by single insertions of the $1/m_Q$ and $1/m_Q^2$ potentials using Rayleigh-Schr\"odinger perturbation theory and contains both SD and LD contributions. The SD part is state-independent and contains a UV divergence, which constitutes a renormalization of the vector or pseudoscalar NRQCD current, while all $n$ dependence in eq.~\eqref{eq:softexpandforWFatoriginCS} is LD.

Motivated by the form of $\delta \psi_{s,n}^{\rm Soft}(0)$ in the $n\to\infty$ limit, we therefore propose a new \textit{soft-improved matching} (SIM) scheme,
\begin{equation}
\psi_{s,n}^{\msbar}(0)=\psi_{s,n}^{\sims}(0)\left[1+\frac{\alpha_s^2(\mu_R)}{2}\delta Z^{\sims}_{s}\right]\,,\label{eq:SIMdef}
\end{equation}
where the finite renormalization counterterm is
\begin{eqnarray}
\delta Z^{\sims}_{s}&=&C_F^2\left(\dfrac{L_\Lambda}{2s+1}+\dfrac{5}{3}+\dfrac{7}{3(2s+1)}\right)\nonumber\\
&&+C_FC_A\left(\dfrac{L_{\Lambda}}{2}+\dfrac{5}{4}\right)\,,\label{eq:softimprovedCT}
\end{eqnarray}
the logarithm $L_{\Lambda}$ is defined as $L_{\Lambda} \equiv \log\left(\mu_{\Lambda}^2/m_Q^2\right)$, and $\psi_{s,n}^{\msbar}(0)$ denotes the renormalized wavefunction at the origin in the $\msbar$ scheme. The $n$-dependent soft, Coulombic, and NP contributions are absorbed into the LD wavefunction at the origin, $\psi_{s,n}^{\sims}(0)$. Since $\alpha_s$ is evaluated at the same renormalization scale $\mu_R$ as the SDCs in the SIM scheme, the $\mu_\Lambda$ dependence cancels between $\delta Z_s^{\sims}$ and the SDCs. In other words, we introduce the renormalization of the LDME $\langle \mathcal{O} \rangle$ via
\begin{equation}
\langle \mathcal{O}_{\mathrm{bare}} \rangle=\langle \mathcal{O}_{\msbar} \rangle Z_s^{\msbar}=\langle \mathcal{O}_{\sims} \rangle Z_s^{\sims}\,,
\end{equation}
where
\begin{eqnarray}
Z_s^{\msbar}&=&1+\alpha_s^2(\mu_R)\left(\frac{C_F^2}{2s+1}+\frac{C_FC_A}{2}\right)\nonumber\\
&&\times\left(\frac{1}{\bar{\epsilon}}+\log{\left(\frac{\mu_R^2}{\mu_\Lambda^2}\right)}\right)\,,\nonumber\\
Z_s^{\sims}&=&Z_s^{\msbar}+\alpha_s^2(\mu_R)\delta Z_s^{\sims}\,.
\end{eqnarray}
In the above equations, we have defined $1/\bar{\epsilon}=1/\epsilon+\log{\left(4\pi\right)}-\gamma_E$, with $\gamma_E$ denoting the Euler-Mascheroni constant.

We point out here that the important point of the SIM scheme is not to cancel the $\mu_{\Lambda}$ dependence, thereby eliminating the need to perform the RG evolution of the LDMEs at NNLO. Instead, there is some freedom in whether to retain $L_\Lambda$ in $\delta Z^{\sims}_s$, as shown in eq.~\eqref{eq:softimprovedCT}, or simply to drop it. In the latter case, the LDMEs still need to be evolved with respect to the NRQCD scale $\mu_{\Lambda}$. The crucial point of the SIM scheme is that the right-hand side of eq.~\eqref{eq:SIMdef} gives rise to additional non-zero finite, non-logarithmic contributions at $\mathcal{O}(\alpha_s^2)$, which are expected to cancel against the large negative constant terms in the SDCs, while leaving the LO and NLO results unchanged. Although the SIM scheme is merely a reorganization of the perturbative calculation, it provides a scheme in which the $\mathcal{O}(\alpha_s^2)$ coefficients are stabilized through the cancellation between the SDCs and $\delta Z_s^{\sims}$. For the $\eta_c\to\gamma\gamma$ example, after incorporating the finite-renormalization contribution from $\delta Z_s^{\sims}$, the NNLO result in eq.~\eqref{eq:etac2aaNNLOnoSoft} becomes
\begin{equation}
\frac{\Gamma_{\mathrm{NNLO}}(\eta_c\to\gamma\gamma)}{\Gamma_{\mathrm{LO}}(\eta_c\to\gamma\gamma)}=1-1.07\alpha_s+\left[\underbrace{-11.16}_{\mathrm{SDC}}+\underbrace{12.11}_{\mathrm{LDME}}\right]\alpha_s^2\,,\label{eq:etac2aaNNLO}
\end{equation}
where the coefficients $-11.16$ and $+12.11$ arise from the original SDC contribution and the new finite LDME renormalization contribution [cf.~eq.~\eqref{eq:softimprovedCT}], respectively. These two contributions largely cancel, yielding a much smaller positive $\alpha_s^2$ coefficient of $+0.95$. As a result, the perturbative convergence is significantly improved, with no rapid growth of the perturbative coefficients and an alternating-sign behavior of the series. More examples can be found in the Supplemental Material~\ref{sec:SM}.

We now assess the perturbative convergence of nine physical observables that have been measured experimentally. We restrict ourselves to NNLO QCD corrections and to the LP contribution in the $v^2$ expansion. For completeness, we note that additional corrections are available for some observables. For instance, the decay $\Upsilon(1S)\to e^+e^-$ has been studied including three-loop QCD radiative corrections~\cite{Beneke:2014qea,Feng:2022vvk}, and relativistic corrections are known for a variety of processes as well. Their implementation, however, goes beyond the scope of this work. We denote by ``NNLO$_{\sims}$" and ``NNLO$_{\msbar}$" the NNLO predictions in the SIM and $\msbar$ schemes, respectively. Furthermore, we focus only on the ground states with principal quantum number $n=1$. We use the charm-quark mass $m_c=1.5$ GeV, the bottom-quark mass $m_b=4.75$ GeV, the electromagnetic coupling $\alpha=1/137.036$, and set the NRQCD factorization scale $\mu_{\Lambda}$ equal to the heavy-quark mass $m_Q$ (the latter being relevant only for the NNLO$_{\msbar}$ results). The widths of the constituent heavy quarks are neglected. The RG evolution of $\alpha_s(\mu_R)$ is implemented using the five-loop QCD beta function~\cite{Baikov:2016tgj,Herzog:2017ohr}, based on an extension of the \texttt{HOPPET} program~\cite{Salam:2008qg} following refs.~\cite{Capatti:2025khs,AbdulHameed:2026utd}. The initial condition is $\alpha_s=0.118$ at $\mu_R=m_Z=91.1876$ GeV. Four-loop decoupling relations for $\alpha_s$~\cite{Chetyrkin:1997un,Schroder:2005hy,Chetyrkin:2005ia,Gerlach:2018hen}, as well as the corresponding loop-order decoupling for SDCs, are employed, with on-shell mass thresholds for the top, bottom, and charm quarks set to $m_t = 172.56~\mathrm{GeV}$, $m_b$, and $m_c$, respectively. In the following, we present only renormalization-scale uncertainties, since other sources of uncertainty, such as parametric ones, are largely independent of the perturbative $\alpha_s$ order. A more detailed assessment of the error budgets for each observable is beyond the scope of this letter.

The first observable we examine here is the \texttt{BaBar} measurement~\cite{BaBar:2010siw} of the absolute transition form-factor ratio for $\gamma\gamma^* \to \eta_c$, $|F(Q^2)/F(0)|$, where $Q^2$ denotes the virtuality of the initial space-like photon $\gamma^*$ and $F(Q^2)$ is the electromagnetic transition form factor. As a ratio, this observable has the advantage of being independent of global normalization factors, which may suffer from large uncertainties, such as the value of the LDME. The first NNLO QCD calculation in the $\msbar$ scheme within NRQCD was reported in ref.~\cite{Feng:2015uha}. That study casts strong doubt on the applicability of the NRQCD approach to $|F(Q^2)/F(0)|$, as the substantial negative $\alpha_s^2$ corrections spoil the good agreement between the NLO prediction and the \texttt{BaBar} data. Refs.~\cite{Wang:2018lry,Babiarz:2025agk} revisit the NNLO calculation by choosing a different renormalization scale than $\mu_R=\sqrt{m_c^2+Q^2}$ used in ref.~\cite{Feng:2015uha}. In particular, ref.~\cite{Babiarz:2025agk} points out that a scale choice of $\mu_R=2m_c$, motivated by the dominant Feynman diagrams, could bring the NNLO QCD prediction closer to the experimental data, although the scale uncertainty associated with $\mu_R$ remains much larger than at NLO. A smaller value of $\mu_\Lambda=1~\mathrm{GeV}$ helps to reduce the size of the $\alpha_s^2$ corrections. From eqs.~\eqref{eq:SIMdef} and \eqref{eq:softimprovedCT}, this can be understood as the fact that lowering $\mu_\Lambda$ effectively shifts part of the soft contributions from the wavefunction at the origin to the SDCs in the $\msbar$ scheme.

\begin{figure}[hbt!]
\includegraphics[width=0.95\columnwidth,draft=false]{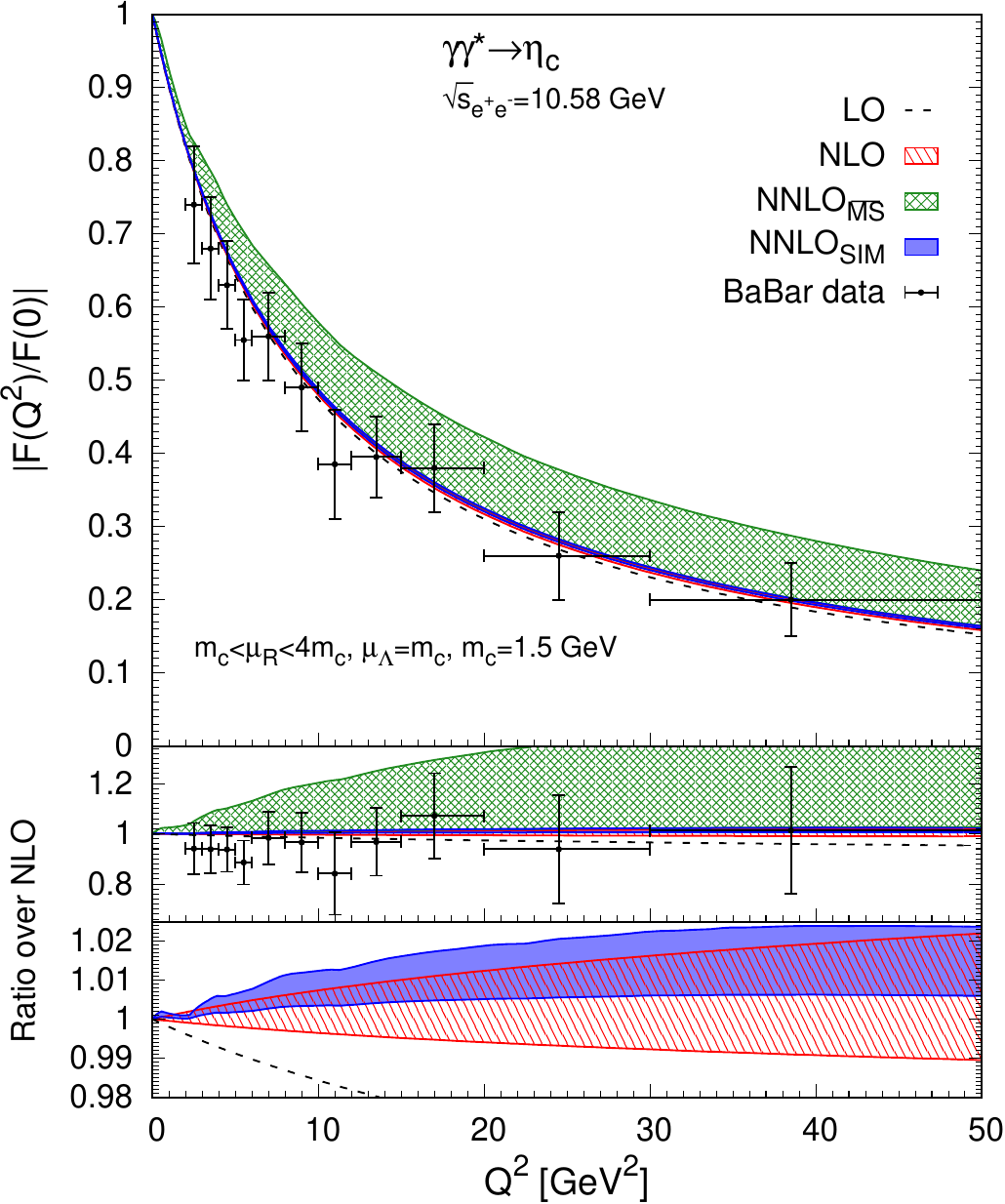}
\caption{The absolute transition form-factor ratio $|F(Q^2)/F(0)|$ for $\gamma\gamma^* \to \eta_c$ as a function of the photon virtuality $Q^2$ is shown at LO (dashed), NLO (red hatched), NNLO$_{\msbar}$ (green hatched), and NNLO$_{\sims}$ (blue band). Experimental data from the {\tt BaBar} experiment are taken from ref.~\cite{BaBar:2010siw}. The lower panels show ratios normalized to the central NLO value. In the bottom panel, both NNLO$_{\msbar}$ and the {\tt BaBar} data are not shown.}
\label{figaastar2etac}
\end{figure}

Our NNLO prediction in the SIM scheme is shown as a blue band in fig.~\ref{figaastar2etac}, together with the LO (dashed), NLO (red hatched), and NNLO$_{\msbar}$ (green hatched) results. The bands represent renormalization-scale uncertainties obtained by varying $\mu_R$ from $m_c$ to $4m_c$. The perturbative convergence improves significantly from NLO to NNLO$_{\sims}$: the NNLO$_{\sims}$ band overlaps with the NLO band, while the scale uncertainty is reduced by a factor of $1.5$. By contrast, the NNLO$_{\msbar}$ result falls outside the NLO band and exhibits substantially larger scale dependence. Overall, the NNLO$_{\sims}$ result shows markedly reduced sensitivity to the renormalization scale compared to NNLO$_{\msbar}$ and is in good agreement with the \texttt{BaBar} measurement.

With this encouraging result, we now investigate eight other observables that are known at NNLO in the $\msbar$ scheme and for which experimental measurements are available. In order to facilitate comparisons with experiment, we fix the normalizations of the LDMEs, which are unfortunately not precisely known. In addition, rigorously speaking, the renormalized LDMEs differ between matching schemes, such that physical observables obtained from the product of the SDCs and LDMEs are independent of the scheme choice. Nevertheless, in the following, we use the same LDME values when studying NNLO QCD perturbative convergence in quarkonium processes. For illustration, we take $\langle \mathcal{O}^{\eta_c}(^1S_0^{[1]})\rangle=0.318~\mathrm{GeV}^{3}$ and $\langle \mathcal{O}^{J/\psi}(^3S_1^{[1]})\rangle=1.36~\mathrm{GeV}^{3}$ for charmonium states, obtained by tuning our central NNLO$_{\sims}$ predictions for $\Gamma(\eta_c\to\gamma\gamma)$ and $\Gamma(J/\psi\to e^+e^-)$ to the central values of the \texttt{BESIII} measurement~\cite{BESIII:2025vdn} and the 2026 Particle Data Group (\texttt{PDG}) value~\cite{ParticleDataGroup:2026aaa}, respectively. For bottomonium states, we take $\langle\mathcal{O}^{\eta_b}(^1S_0^{[1]})\rangle = 3.093~\mathrm{GeV}^{3}$ and $\langle \mathcal{O}^{\Upsilon(1S)}(^3S_1^{[1]})\rangle = 9.28~\mathrm{GeV}^{3}$ from ref.~\cite{Eichten:1995ch}, using the Buchm\"uller-Tye QCD potential~\cite{Buchmuller:1980su}. We stress that the data-theory comparisons presented here should be interpreted with caution, as higher-order QCD radiative and relativistic corrections, as well as parametric uncertainties, are not included. Nevertheless, it is natural to ask whether there exist regions of parameter space that yield improved agreement within the present setup, potentially providing an important test of the NRQCD factorization framework at NNLO.

\begin{table*}[htpb!]
\centering
\tabcolsep=3.5mm
\renewcommand*{\arraystretch}{1.3}
\vspace{0.2cm}
\begin{tabular}{|c|c|c|c|c|c|} \hline
 \multirow{2}{*}{Observable} & \multirow{2}{*}{Experiment} & \multicolumn{4}{c|}{Theory} \\\cline{3-6}
 & & LO & NLO & NNLO$_{\msbar}$ & NNLO$_{\sims}$ \\ \hline
$\Gamma(\eta_c\to gg)$ [MeV] &  $30.0\pm 0.5$~\cite{ParticleDataGroup:2026aaa} &  $12.7^{+11.5}_{-4.6}$& $23.5^{+13.8}_{-6.9}$ & $19.4^{+0.0}_{-20.0}$ & $29.2^{+6.1}_{-6.2}$\\
$\Gamma(\eta_c\to \gamma \gamma)$ [keV] &  $6.4\pm0.6$~\cite{ParticleDataGroup:2026aaa} &  $9.3$ & $6.8^{+0.5}_{-1.0}$ & $-0.5_{-6.5}^{+2.8}$ & $6.8^{+0.1}_{-0.1}$ \\
$\Gamma(\eta_b\to gg)$ [MeV] &  $10^{+5}_{-4}$~\cite{ParticleDataGroup:2026aaa} &  $6.2^{+2.7}_{-1.6}$ & $9.7^{+2.0}_{-1.7}$ & $8.3_{-2.2}^{+0.1}$ & $10.7^{+0.4}_{-1.0}$ \\
$\Gamma(J/\psi\to \gamma \gamma \gamma)$ [eV] &  $1.1\pm 0.2$~\cite{ParticleDataGroup:2026aaa} &  $5.3$ & $-0.1_{-2.1}^{+1.1}$ & $-2.3_{-1.7}^{+1.1}$ & $0.8^{+1.2}_{-0.1}$ \\
$\Gamma(J/\psi\to e^+e^-)$ [keV] &  $5.5\pm 0.1$~\cite{ParticleDataGroup:2026aaa} &  $10.0$ & $5.7_{-1.6}^{+0.9}$ & $-0.5_{-5.3}^{+2.5}$ & $5.5^{+0.3}_{-0.0}$ \\
$\Gamma(\Upsilon(1S)\to e^+e^-)$ [keV] &  $1.3\pm 0.1$~\cite{ParticleDataGroup:2026aaa} &  $1.7$ & $1.2_{-0.1}^{+0.1}$ & $0.7_{-0.2}^{+0.2}$ & $1.2^{+0.0}_{-0.0}$ \\
$\sigma(e^+e^-\to J/\psi\eta_c)$ [fb] &  $17.6^{+3.2}_{-3.5}$~\cite{BaBar:2005nic} &  $3.5^{+2.1}_{-1.0}$ & $7.3_{-1.9}^{+3.4}$ & $9.6_{-1.8}^{+2.1}$ & $12.9^{+7.3}_{-3.5}$ \\
$\sigma(e^+e^-\to J/\psi J/\psi)$ [fb] &  $<9.1$~\cite{Belle:2004abn} &  $7.5$ & $1.9_{-1.5}^{+0.9}$ & $-4.3_{-3.6}^{+2.0}$ & $1.8^{+0.2}_{-0.0}$ \\\hline
\end{tabular}
\caption{Comparison of theoretical predictions at different perturbative orders in $\alpha_s$ with experimental measurements~\cite{ParticleDataGroup:2026aaa,BaBar:2005nic,Belle:2004abn}. The theoretical uncertainties arise from renormalization-scale variations around the central scale $\mu_{R,0}$, with $\mu_{R,0}=2m_Q$ for the first six decay widths and $\mu_{R,0}=\sqrt{s}/2$ for the last two cross sections at $\sqrt{s}=10.58~\mathrm{GeV}$.\label{tab:widthxs}
}
\end{table*}

The theoretical predictions at different perturbative orders in $\alpha_s$ are presented in tab.~\ref{tab:widthxs} for six decay processes, $\eta_c \to gg$, $\eta_c \to \gamma\gamma$, $\eta_b \to gg$, $J/\psi \to \gamma\gamma\gamma$, $J/\psi \to e^+e^-$, and $\Upsilon(1S) \to e^+e^-$, as well as for two exclusive double-charmonium production processes, $e^+e^- \to J/\psi \eta_c$ and $e^+e^- \to J/\psi J/\psi$, in $e^+e^-$ collisions at $\sqrt{s} = 10.58$ GeV. The NNLO$_{\msbar}$ results are taken from refs.~\cite{Feng:2017hlu,Czarnecki:2001zc,Feng:2015uha,Abreu:2022cco,Czarnecki:1997vz,Beneke:1997jm,Beneke:2014qea,Feng:2022vvk,Zeng:2026ois,Feng:2019zmt,Huang:2022dfw,Li:2025mng,Chen:2025qgy,Sang:2023liy,Huang:2023pmn}. The central renormalization scale is chosen as $\mu_{R,0} = 2m_Q$ for decay processes and $\mu_{R,0} = \sqrt{s}/2$ for production processes. The theoretical uncertainties are estimated by varying $\mu_R$ around $\mu_{R,0}$ by a factor of two.

These processes involve rather different physics. The SDCs of the decay widths involve a single scale $m_Q$, whereas the SD cross sections for double-charmonium production involve two scales, $\sqrt{s}$ and $m_Q$. All processes except $\eta_Q\to gg$ ($Q=c,b$) are exclusive and thus receive only virtual corrections in perturbative calculations. Due to large negative $\alpha_s^2$ coefficients, the NNLO$_{\msbar}$ predictions for $\eta_c\to\gamma\gamma$, $J/\psi\to \gamma\gamma\gamma$, $J/\psi\to e^+e^-$, and $e^+e^-\to J/\psi J/\psi$ become negative at the central scale $\mu_R=\mu_{R,0}$, and $\Gamma(\eta_c\to gg)$ also becomes negative when $\mu_R=m_c$. On the other hand, all NNLO$_{\sims}$ results are positive definite. The fractional renormalization-scale uncertainties are significantly reduced from NNLO$_{\msbar}$ to NNLO$_{\sims}$, except for $\sigma(e^+e^-\to J/\psi\eta_c)$. The origin of the different behavior observed for $\sigma(e^+e^-\to J/\psi\eta_c)$ remains unclear. One possible explanation is the presence of large logarithms, $\log(s/m_c^2)$, which have been discussed extensively in the literature~\cite{Jia:2010fw,Bodwin:2014dqa}. These logarithmic contributions are unrelated to the issue addressed in this letter and would likely require resummation in order to improve the theoretical prediction. Another interesting observation concerns $\Gamma(J/\psi\to \gamma\gamma\gamma)$, whose NLO corrections already render the central value of the partial width negative. Inspecting its perturbative coefficients,
\begin{equation}
\frac{\Gamma_{\mathrm{NNLO}}(J/\psi\to\gamma\gamma\gamma)}{\Gamma_{\mathrm{LO}}(J/\psi\to\gamma\gamma\gamma)}=1-4.02\alpha_s+\left[\underbrace{-6.89}_{\mathrm{SDC}}+\underbrace{9.35}_{\mathrm{LDME}}\right]\alpha_s^2\,\label{eq:Jpsi2aaaNNLO}
\end{equation}
with $\mu_R=2m_c$, one finds a large negative $\alpha_s$ coefficient, $-4.02$, while $\alpha_s(2m_c)\approx 0.253$.
The inclusion of the finite contributions from $\delta Z_s^{\sims}$ renders the partial width positive at NNLO, while the NNLO$_{\msbar}$ contribution becomes even more negative. We emphasize that, despite their very different physical nature, both $\Gamma(J/\psi\to\gamma\gamma\gamma)$ and $\Gamma(J/\psi\to e^+e^-)$ agree with the \texttt{PDG} 2026 values~\cite{ParticleDataGroup:2026aaa} at NNLO using a common LDME, which can be even more clearly seen from their ratio in the Supplemental Material~\ref{sec:SM} that is independent of the LDME. The same conclusion applies to $\Gamma(\eta_c\to gg)$ and $\Gamma(\eta_c\to \gamma\gamma)$. The renormalization-scale dependence and more extensive theory-data comparisons are provided in the Supplemental Material~\ref{sec:SM}.

Our study shows that the large NNLO corrections to the $\msbar$ SDCs are an artifact of minimal subtraction, which splits a single potential-region contribution into equal and opposite contributions to the SDC and the wavefunction at the origin. Once the $n$-independent soft finite renormalization $\delta Z_s^{\sims}$ is combined with the SDC, the genuinely hard correction is small. Since $\delta Z_s^{\sims}$ is purely perturbative and the cancellation persists without invoking a $\beta_0$-driven renormalon, the effect is therefore SD in origin and does not rely on an IR cutoff associated with a confining QCD potential~\cite{Chung:2020zqc}. Our approach also differs from the weak-coupling approach of refs.~\cite{Kiyo:2010jm,Beneke:2014qea}. Unlike the weak-coupling analyses of refs.~\cite{Kiyo:2010jm,Beneke:2014qea}, we do not assume that the soft and ultrasoft scales are perturbative. In the SIM scheme, only $\delta Z_s^{\sims}$ is treated perturbatively. This trades systematic improvability and the resummation of the hard-to-soft logarithm for applicability to charmonium, where $m_Qv \sim 700$-$800$ MeV and $m_Qv^2\sim \Lambda_{\rm QCD}$. At NNLO, in the limit $m_Qv\gg \Lambda_{\rm QCD}$, our approach reduces to the weak-coupling approach.

In summary, we investigate in this letter the issue of perturbative instability in NNLO QCD corrections to quarkonium production and decay processes. We point out that this instability originates from the generic presence of rapidly growing negative $\alpha_s^2$ coefficients in the traditional $\msbar$ scheme for the renormalization of LDMEs. Such a renormalization scheme is not optimal, since these large negative $\alpha_s^2$ coefficients are expected to cancel against large positive contributions from the SD soft (non-Coulombic) components of the wavefunctions at the origin. We therefore introduce a new SIM scheme in this letter. Including the finite soft renormalization contribution $\delta Z_s^{\sims}$ [cf. eq.~\eqref{eq:softimprovedCT}] in the NNLO perturbative calculations significantly reduces the magnitude of the $\alpha_s^2$ coefficients. We have examined nine physical observables that are known at NNLO accuracy in the literature and have been measured experimentally. The SIM scheme significantly improves the NNLO theoretical predictions: all partial widths and cross sections become positive, and the renormalization-scale uncertainties are substantially reduced, with the exception of $\sigma(e^+e^-\to J/\psi\eta_c)$, which nonetheless remains affected by large logarithms. Overall, the agreement between theory and data is markedly improved. Finally, the NNLO stability in the SIM scheme enables, for the first time, a consistent determination of the LDMEs from experimental data at this order.  


{Acknowledgments} -- This work is supported by the grants from the ERC (grant 101041109 `BOSON') and the French ANR (grant ANR-20-CE31-0015 `PrecisOnium'). Views and opinions expressed are however those of the authors only and do not necessarily reflect those of the European Union or the European Research Council Executive Agency. Neither the European Union nor the granting authority can be held responsible for them.

\bibliographystyle{utphys}

\bibliography{reference}

\onecolumngrid

\newpage
\appendix

\section{Supplemental Material}\label{sec:SM}

In this Supplemental Material, we present further details on perturbative coefficients, scale dependence, and theory-data comparisons for the quarkonium processes under reappraisal.

Unless stated otherwise, we use the following input parameters:
\begin{equation}
\begin{aligned}
m_c&=1.5~\mathrm{GeV}\,,\quad m_b=4.75~\mathrm{GeV}\,,\quad m_t=172.56~\mathrm{GeV}\,,\quad \mu_{\Lambda}=m_Q\,,\quad \alpha^{-1}=137.036\,,\\
m_W&=80.3625~\mathrm{GeV}\,,\quad m_Z=91.1876~\mathrm{GeV}\,,\quad m_H=125.20~\mathrm{GeV}\,.\label{eq:inputparameters}
\end{aligned}
\end{equation}
The decay widths of the constituent heavy quarks are neglected. We focus only on the ground states with principal quantum number $n=1$.

For the perturbative coefficients, we fix the number of light-flavor quarks to $n_f=3,4,5$ for charmonium, bottomonium, and toponium, respectively, independent of the renormalization scale $\mu_R$. The appropriate decoupling relations are applied in the short-distance coefficients (SDCs).

On the other hand, to illustrate the scale dependence of physical observables and the theory-data comparison, we use the five-loop QCD evolution of $\alpha_s(\mu_R)$ with the initial condition $\alpha_s(m_Z)=0.118$, together with four-loop decoupling relations for $\alpha_s$, where the on-shell mass thresholds for the top, bottom, and charm quarks are set to $m_t=172.56$ GeV, $m_b=4.75$ GeV, and $m_c=1.5$ GeV, respectively. The corresponding loop-order decoupling for the SDCs is also implemented. For comparison with experimental measurements, we take
\begin{equation}
\langle \mathcal{O}^{\eta_c}(^1S_0^{[1]})\rangle=0.318~\mathrm{GeV}^{3}\,,\quad
\langle \mathcal{O}^{J/\psi}(^3S_1^{[1]})\rangle=1.36~\mathrm{GeV}^{3}\,
\end{equation}
for charmonium states, obtained by tuning our central NNLO$_{\sims}$ predictions for $\Gamma(\eta_c\to\gamma\gamma)$ and $\Gamma(J/\psi\to e^+e^-)$ to the central values of the \texttt{BESIII} measurement~\cite{BESIII:2025vdn} and the \texttt{PDG} 2026 value~\cite{ParticleDataGroup:2026aaa}, respectively. For bottomonium states, we take
\begin{equation}
\langle\mathcal{O}^{\eta_b}(^1S_0^{[1]})\rangle = 3.093~\mathrm{GeV}^{3}\,,\quad
\langle \mathcal{O}^{\Upsilon(1S)}(^3S_1^{[1]})\rangle = 9.28~\mathrm{GeV}^{3}\,
\end{equation}
from ref.~\cite{Eichten:1995ch}, based on the Buchm\"uller-Tye QCD potential~\cite{Buchmuller:1980su}.

Finally, we define $V_Q$ as a shorthand for the ground-state spin-triplet $S$-wave quarkonia: $V_c=J/\psi$, $V_b=\Upsilon(1S)$, and $V_t$ the $1S$ spin-triplet toponium state.

\subsection{Perturbative Coefficients and Scale Dependence}

\subsubsection{$\eta_Q\to \gamma\gamma$}

\begin{figure}[hbt!]
\subfloat[]{\includegraphics[width=0.49\columnwidth,draft=false]{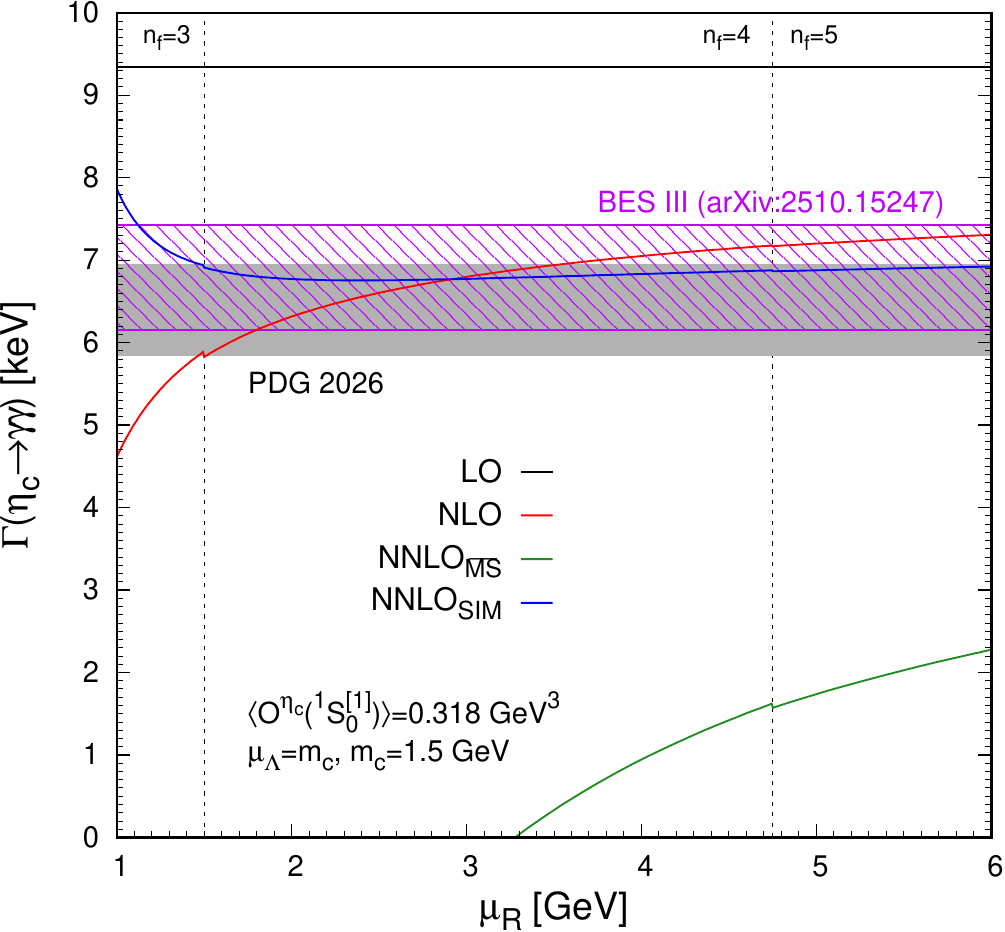}\label{fig:etaQdecay-a}}\hfill
\subfloat[]{\includegraphics[width=0.49\columnwidth,draft=false]{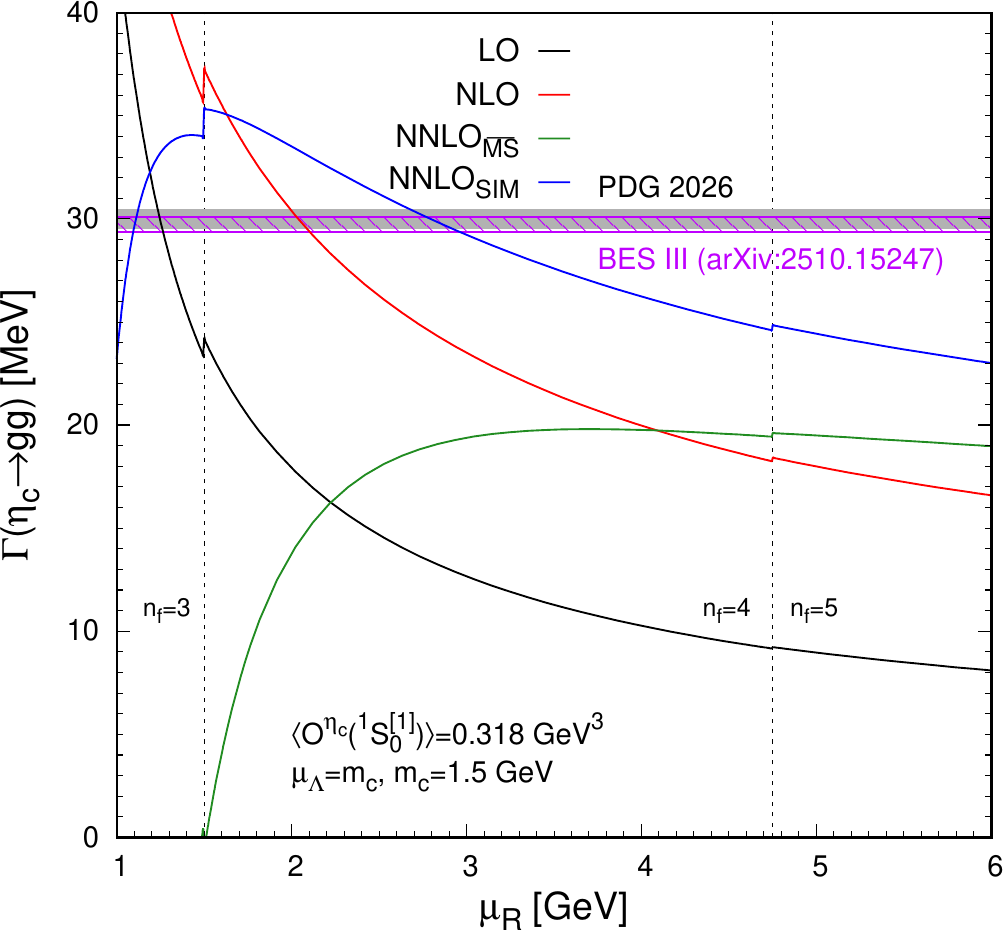}\label{fig:etaQdecay-b}}\\
\subfloat[]{\includegraphics[width=0.49\columnwidth,draft=false]{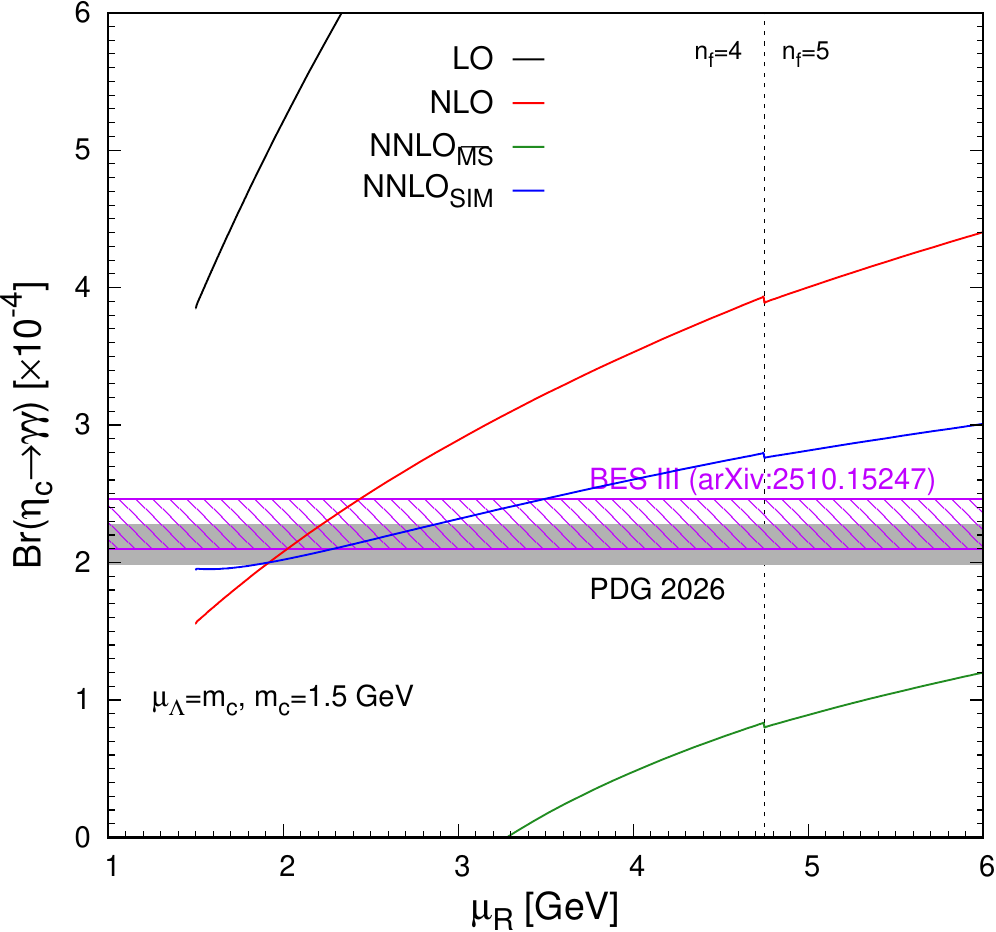}\label{fig:etaQdecay-c}}\hfill
\subfloat[]{\includegraphics[width=0.49\columnwidth,draft=false]{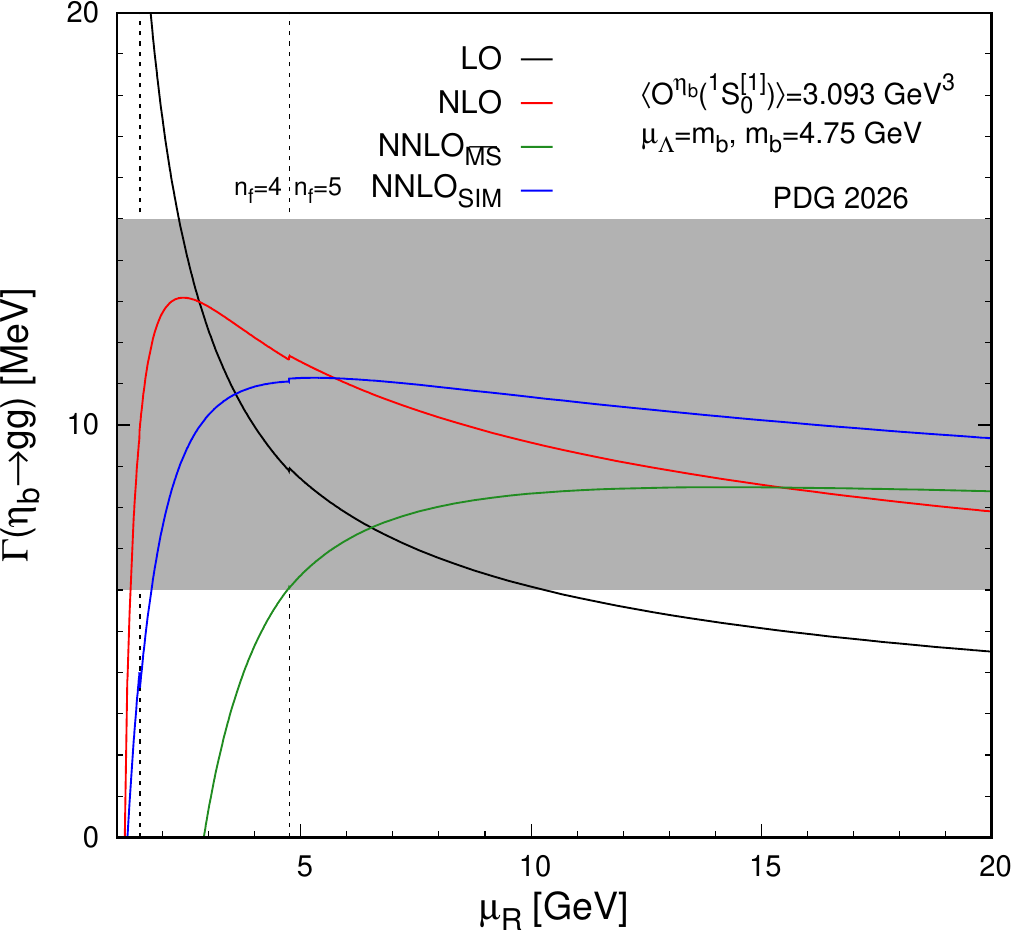}\label{fig:etaQdecay-d}}
\caption{Renormalization-scale dependence of the partial decay widths $\eta_c\to\gamma\gamma$ (upper left), $\eta_c\to gg$ (upper right), $\eta_b\to gg$ (lower right), and the branching fraction $\mathrm{Br}(\eta_c\to \gamma\gamma)$ (lower left). \texttt{PDG} (gray band) and \texttt{BESIII} (purple hatched) data are taken from refs.~\cite{ParticleDataGroup:2026aaa} and~\cite{BESIII:2025vdn}, respectively.}
\label{fig:etaQdecay}
\end{figure}

NNLO QCD corrections to $\Gamma(\eta_Q\to \gamma\gamma)$ in NRQCD with the $\msbar$-renormalized long-distance matrix element (LDME) are known from refs.~\cite{Czarnecki:2001zc,Feng:2015uha,Abreu:2022cco}. The perturbative $\alpha_s$ coefficients up to NNLO read
\begin{equation}
\left.\frac{\Gamma_{\mathrm{NNLO}}(\eta_Q\to\gamma\gamma)}{\Gamma_{\mathrm{LO}}(\eta_Q\to\gamma\gamma)}\right|_{\mu_R=m_Q}=1-1.07\alpha_s+\left[\left(\begin{array}{c}-11.16_{\mathrm{SDC},n_f=3}\\
-10.40_{\mathrm{SDC},n_f=4}\\
-11.19_{\mathrm{SDC},n_f=5}
\end{array}\right)+12.11_{\mathrm{LDME}}\right]\alpha_s^2\,,
\end{equation}
where the term labeled ``LDME" is set to zero for NNLO$_{\overline{\mathrm{MS}}}$ and to $+12.11$ for NNLO$_{\mathrm{SIM}}$. The same convention will be applied to the other processes discussed in this Supplemental Material. The renormalization-scale dependence of the partial decay width $\Gamma(\eta_c\to \gamma\gamma)$ at different perturbative orders is shown in fig.~\ref{fig:etaQdecay-a}. We also compare our theoretical predictions with the \texttt{PDG} 2026~\cite{ParticleDataGroup:2026aaa} and \texttt{BESIII}~\cite{BESIII:2025vdn} data.

\subsubsection{$\eta_Q\to gg$}

NNLO QCD corrections to $\Gamma(\eta_Q\to gg)$ in NRQCD have been computed in ref.~\cite{Feng:2017hlu}, whose $\alpha_s$ expansion reads
\begin{equation}
\left.\frac{\Gamma_{\mathrm{NNLO}}(\eta_Q\to gg)}{\Gamma_{\mathrm{LO}}(\eta_Q\to gg)}\right|_{\mu_R=m_Q}=1+\left(\begin{array}{c}1.54_{n_f=3}\\
1.41_{n_f=4}\\
1.27_{n_f=5}\\
\end{array}\right)\alpha_s+\left[\left(\begin{array}{c}-12.70_{\mathrm{SDC},n_f=3}\\
-13.37_{\mathrm{SDC},n_f=4}\\
-14.06_{\mathrm{SDC},n_f=5}
\end{array}\right)+12.11_{\mathrm{LDME}}\right]\alpha_s^2\,.
\end{equation}
The renormalization-scale dependence of the partial decay widths $\Gamma(\eta_c\to gg)$ and $\Gamma(\eta_b\to gg)$, as well as the branching fraction $\mathrm{Br}(\eta_c\to \gamma\gamma)\approx\Gamma(\eta_c\to \gamma\gamma)/\Gamma(\eta_c\to gg)$, are shown in figs.~\ref{fig:etaQdecay-b}, \ref{fig:etaQdecay-d}, and \ref{fig:etaQdecay-c}, respectively. The experimental data are taken from refs.~\cite{ParticleDataGroup:2026aaa,BESIII:2025vdn}. We note that the branching ratio $\mathrm{Br}(\eta_c\to\gamma\gamma)$ is independent of the normalization of the LDME $\langle \mathcal{O}^{\eta_c}(^1S_0^{[1]})\rangle$.

\subsubsection{$V_Q\to \ell^+\ell^-$}
\begin{figure}[hbt!]
\subfloat[]{\includegraphics[width=0.49\columnwidth,draft=false]{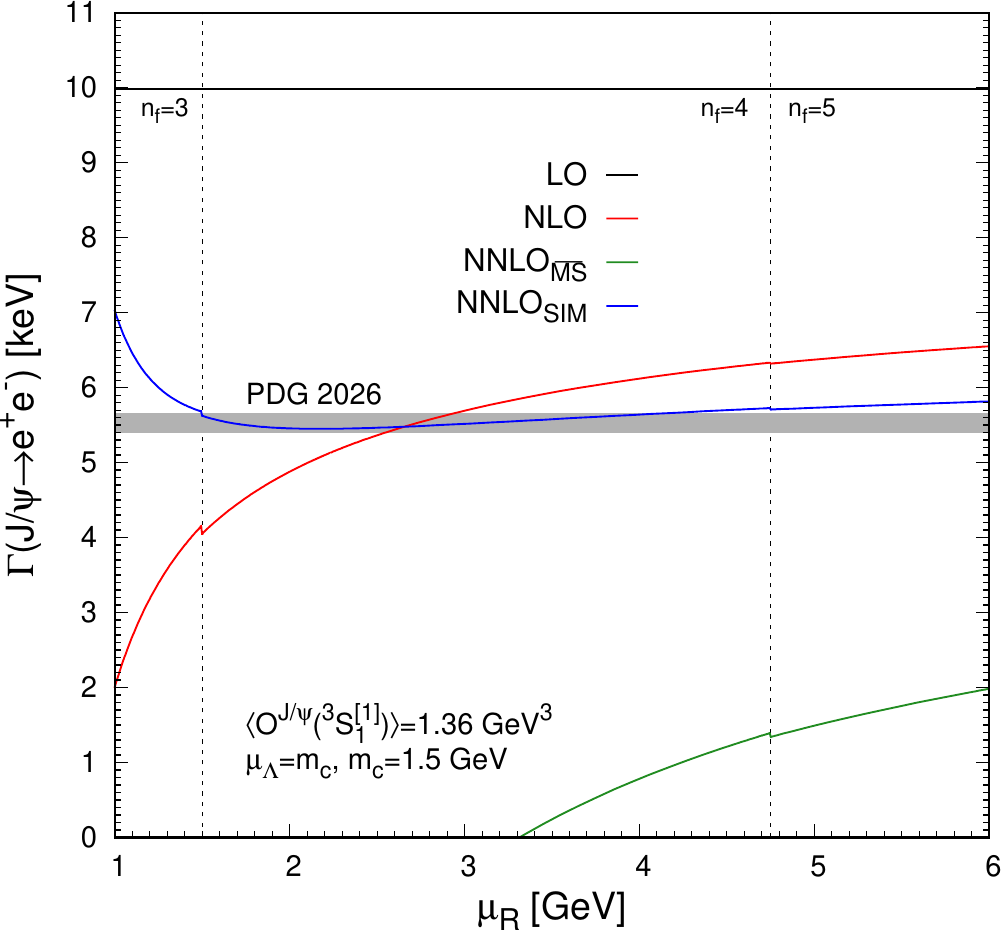}\label{fig:VQdecay-a}}
\hfill
\subfloat[]{\includegraphics[width=0.49\columnwidth,draft=false]{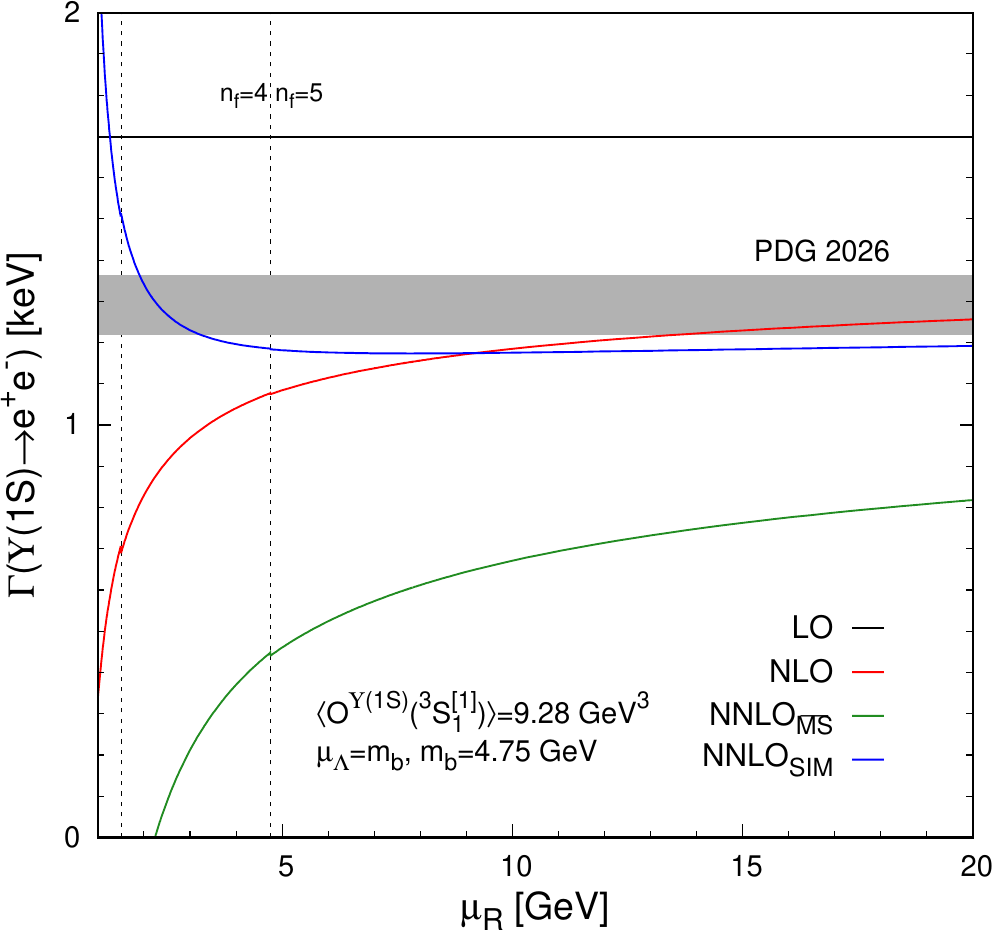}\label{fig:VQdecay-b}}\\
\subfloat[]{\includegraphics[width=0.49\columnwidth,draft=false]{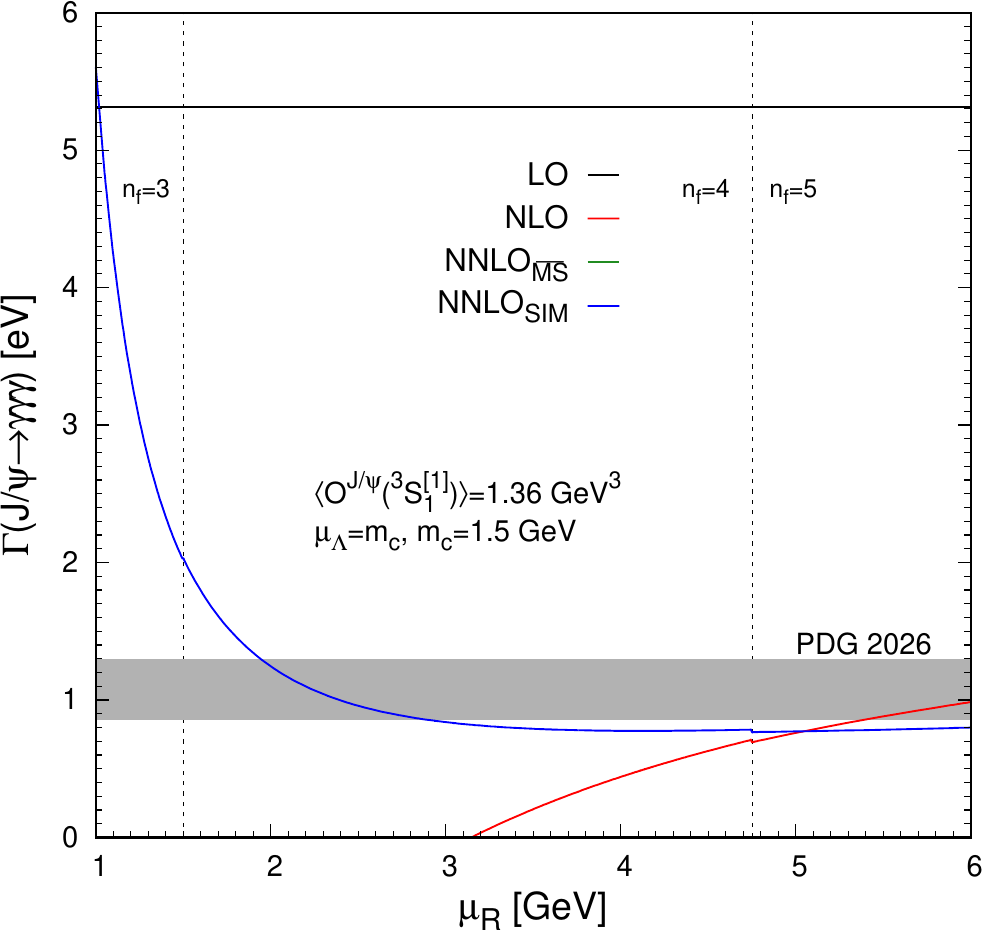}\label{fig:VQdecay-c}}
\subfloat[]
{\includegraphics[width=0.49\columnwidth,draft=false]{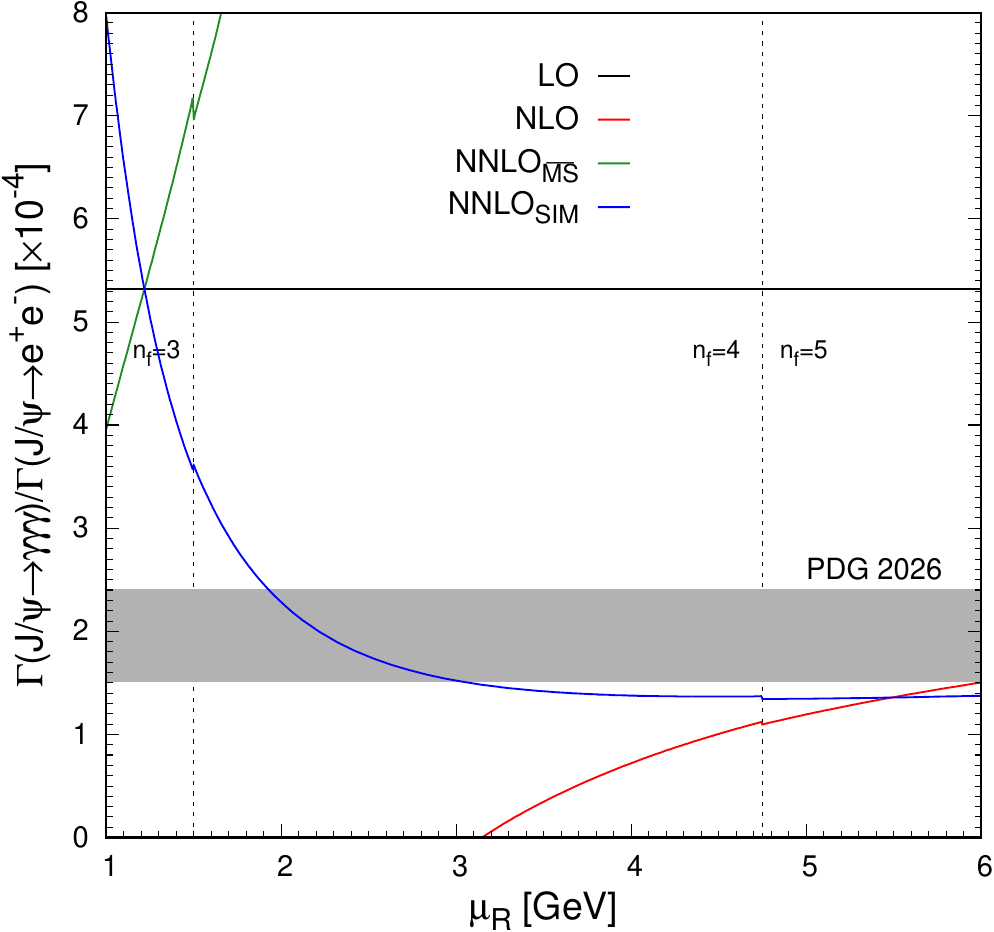}\label{fig:VQdecay-d}}
\caption{Renormalization-scale dependence of the partial decay widths $J/\psi\to e^+e^-$ (upper left), $\Upsilon(1S)\to e^+e^-$ (upper right), $J/\psi\to \gamma \gamma\gamma$ (lower left), and partial width ratio $\Gamma(J/\psi\to \gamma \gamma\gamma)/\Gamma(J/\psi\to e^+e^-)$ (lower right). \texttt{PDG} data (gray band) are taken from ref.~\cite{ParticleDataGroup:2026aaa}. For $J/\psi\to\gamma\gamma\gamma$, the NNLO$_{\msbar}$ result is negative and is therefore not visible.}
\label{fig:VQdecay}
\end{figure}

\vspace*{-1.2mm}
The partial decay width of a vector quarkonium $V_Q$ into a massless lepton pair $\ell^+\ell^-$ through a virtual photon has been computed at NNLO~\cite{Czarnecki:1997vz,Beneke:1997jm,Czarnecki:2001zc} and N$^3$LO~\cite{Beneke:2014qea,Feng:2022vvk} in QCD. In this work, we consider results up to NNLO accuracy. The $\alpha_s$ expansion reads
\vspace*{-1.2mm}
\begin{equation}
\left.\frac{\Gamma_{\mathrm{NNLO}}(V_Q\to\ell^+\ell^-)}{\Gamma_{\mathrm{LO}}(V_Q\to\ell^+\ell^-)}\right|_{\mu_R=m_Q}=1-1.70\alpha_s+\left[\left(\begin{array}{c}-8.06_{\mathrm{SDC},n_f=3}\\
-7.98_{\mathrm{SDC},n_f=4}\\
-7.89_{\mathrm{SDC},n_f=5}
\end{array}\right)+9.35_{\mathrm{LDME}}\right]\alpha_s^2\,.
\end{equation}
The partial decay widths $J/\psi\to e^+e^-$ and $\Upsilon(1S)\to e^+e^-$ are shown in figs.~\ref{fig:VQdecay-a} and \ref{fig:VQdecay-b}, together with a comparison to the \texttt{PDG} 2026 data~\cite{ParticleDataGroup:2026aaa}. 

\subsubsection{$V_Q\to\gamma\gamma\gamma$}

For $V_Q\to \gamma\gamma\gamma$, we take the NNLO QCD calculation from ref.~\cite{Zeng:2026ois}. The corresponding $\alpha_s$ expansion reads
\begin{equation}
\left.\frac{\Gamma_{\mathrm{NNLO}}(V_Q\to\gamma\gamma\gamma)}{\Gamma_{\mathrm{LO}}(V_Q\to\gamma\gamma\gamma)}\right|_{\mu_R=m_Q}=1-4.02\alpha_s+\left[\left(\begin{array}{c}-2.90_{\mathrm{SDC},n_f=3}\\
-4.07_{\mathrm{SDC},n_f=4}\\
-2.07_{\mathrm{SDC},n_f=5}
\end{array}\right)+9.35_{\mathrm{LDME}}\right]\alpha_s^2\,.
\end{equation}
To stay consistent with other results shown in this Supplemental Material, the coefficients are given for $\mu_R=m_Q$, rather than $\mu_R=2m_Q$ as in eq.~\eqref{eq:Jpsi2aaaNNLO}. The results for $\Gamma(J/\psi\to \gamma\gamma\gamma)$ and $\Gamma(J/\psi\to \gamma\gamma\gamma)/\Gamma(J/\psi\to e^+e^-)$ as functions of $\mu_R$, together with a comparison to the \texttt{PDG} value~\cite{ParticleDataGroup:2026aaa}, are shown in figs.~\ref{fig:VQdecay-c} and \ref{fig:VQdecay-d}, respectively. The width ratio is independent of the normalization of the LDME. Since the widths become negative at NLO and NNLO$_{\overline{\mathrm{MS}}}$, ref.~\cite{Zeng:2026ois} proposes using the squared amplitude directly to enforce a positive-definite result. With our improvement, the perturbative expansion in $\alpha_s$ can be consistently performed at the squared-amplitude level up to NNLO.

\subsubsection{$\gamma\gamma^*\to\eta_Q$}

The cross section for $\gamma\gamma^*\to \eta_Q$ in $e^+e^-$ collisions involves two scales: the virtuality $Q^2$ of the space-like $\gamma^*$ and the heavy-quark mass $m_Q$. The electromagnetic form factor $F(Q^2)$ has been computed at NNLO QCD accuracy in refs.~\cite{Feng:2015uha,Babiarz:2025agk}. The perturbative coefficients at $Q=m_Q$ read
\begin{equation}
\left.\frac{d\sigma_{\mathrm{NNLO}}/dQ^2}{d\sigma_{\mathrm{LO}}/dQ^2}\right|_{\mu_R=Q=m_Q}=1-1.05\alpha_s+\left[\left(\begin{array}{c}-11.19_{\mathrm{SDC},n_f=3}\\
-11.21_{\mathrm{SDC},n_f=4}\\
-11.34_{\mathrm{SDC},n_f=5}
\end{array}\right)+12.11_{\mathrm{LDME}}\right]\alpha_s^2\,,
\end{equation}
while those at $Q=7m_Q$ are
\begin{equation}
\left.\frac{d\sigma_{\mathrm{NNLO}}/dQ^2}{d\sigma_{\mathrm{LO}}/dQ^2}\right|_{\mu_R=m_Q,Q=7m_Q}=1-0.57\alpha_s+\left[\left(\begin{array}{c}-11.54_{\mathrm{SDC},n_f=3}\\
-15.13_{\mathrm{SDC},n_f=4}\\
-12.32_{\mathrm{SDC},n_f=5}
\end{array}\right)+12.11_{\mathrm{LDME}}\right]\alpha_s^2\,.
\end{equation}
A comparison between the theoretical prediction for the form-factor ratio $|F(Q^2)/F(0)|$ and the \texttt{BaBar} measurement~\cite{BaBar:2010siw} is shown in fig.~\ref{figaastar2etac}.

\subsubsection{$e^+e^-\to \eta_c\gamma$}

\begin{figure}[hbt!]
\includegraphics[width=0.95\columnwidth,draft=false]{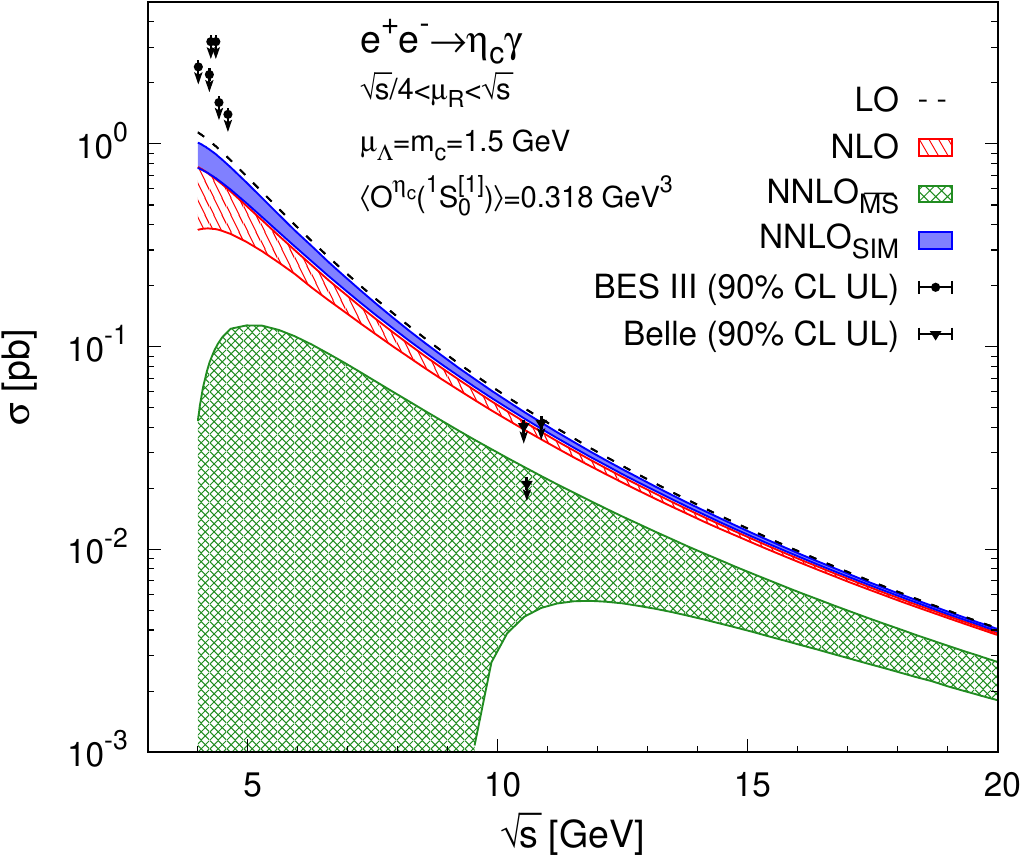}
\caption{The total cross section for $e^+e^-\to \eta_c \gamma$ as a function of $\sqrt{s}$. The $90\%$ CL upper limits from \texttt{BESIII}~\cite{BESIII:2017nty} and \texttt{Belle}~\cite{Belle:2018jqa} are also shown.}
\label{figee2etaca} \vspace*{-0.5cm}
\end{figure}

The NNLO cross section for $e^+e^-\to \eta_c\gamma$ through a time-like virtual photon has been computed in refs.~\cite{Chen:2017pyi,Yu:2020tri,Li:2025pbt}. The $\alpha_s$ expansion at $\sqrt{s}=3m_c$ reads
\begin{equation}
\left.\frac{\sigma_{\mathrm{NNLO}}(e^+e^-\to\eta_c\gamma)}{\sigma_{\mathrm{LO}}(e^+e^-\to\eta_c\gamma)}\right|_{\mu_R=\sqrt{s}/2,\sqrt{s}=3m_c}=1-1.39\alpha_s+\left[
-10.65_{\mathrm{SDC}}
+12.11_{\mathrm{LDME}}\right]\alpha_s^2\,,
\end{equation}
while at $\sqrt{s}=10m_c$ it reads
\begin{equation}
\left.\frac{\sigma_{\mathrm{NNLO}}(e^+e^-\to\eta_c\gamma)}{\sigma_{\mathrm{LO}}(e^+e^-\to\eta_c\gamma)}\right|_{\mu_R=\sqrt{s}/2,\sqrt{s}=10m_c}=1-0.55\alpha_s+\left[
-10.83_{\mathrm{SDC}}+12.11_{\mathrm{LDME}}\right]\alpha_s^2\,.
\end{equation}
$90\%$ confidence level (CL) upper limits from \texttt{BESIII}~\cite{BESIII:2017nty} and \texttt{Belle}~\cite{Belle:2018jqa} are available. A comparison between the theoretical predictions and the experimental upper limits is shown in fig.~\ref{figee2etaca}. We note that the two data points at $\sqrt{s}=10.52$ GeV and $10.58$~GeV lie below the NNLO$_{\mathrm{SIM}}$ predictions, which warrants further investigation.

\subsubsection{$e^+e^-\to J/\psi\eta_c$}

\begin{figure}[hbt!]
\subfloat[]{\includegraphics[width=0.49\columnwidth,draft=false]{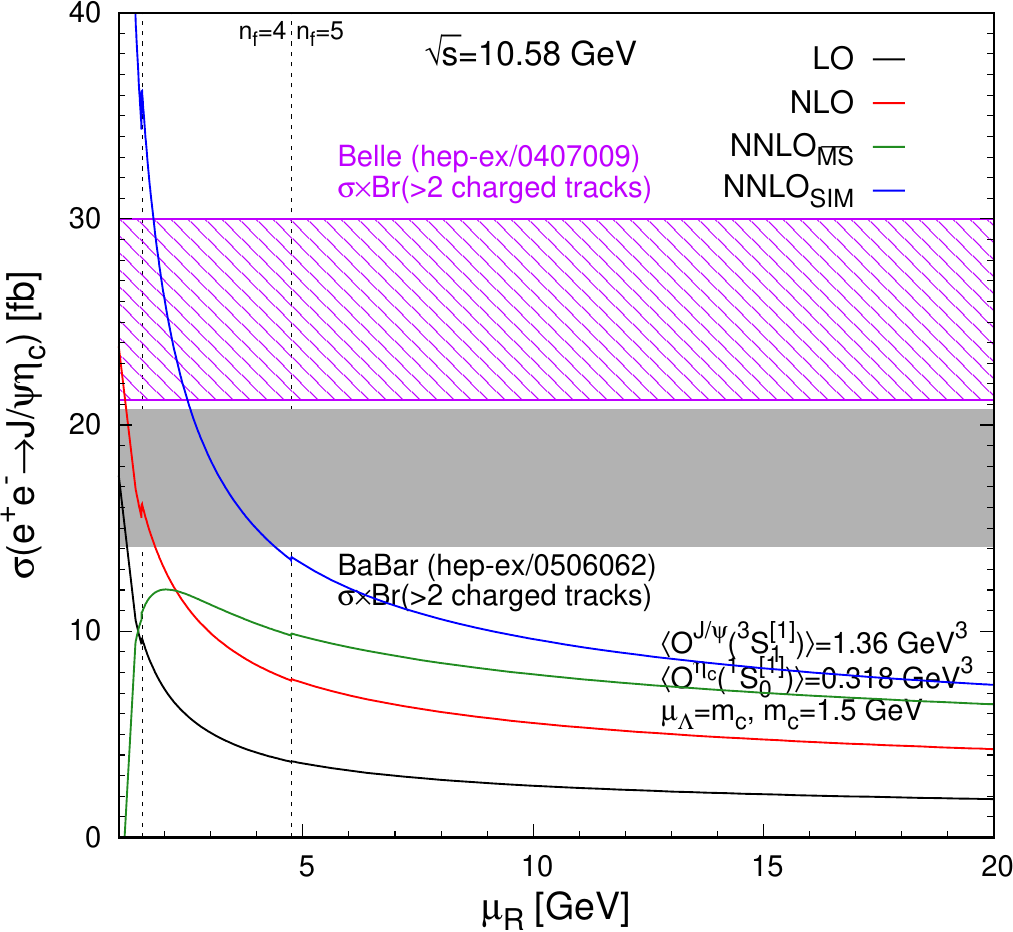}\label{fig:doublecharmonium-a}} \hfill
\subfloat[]{\includegraphics[width=0.49\columnwidth,draft=false]{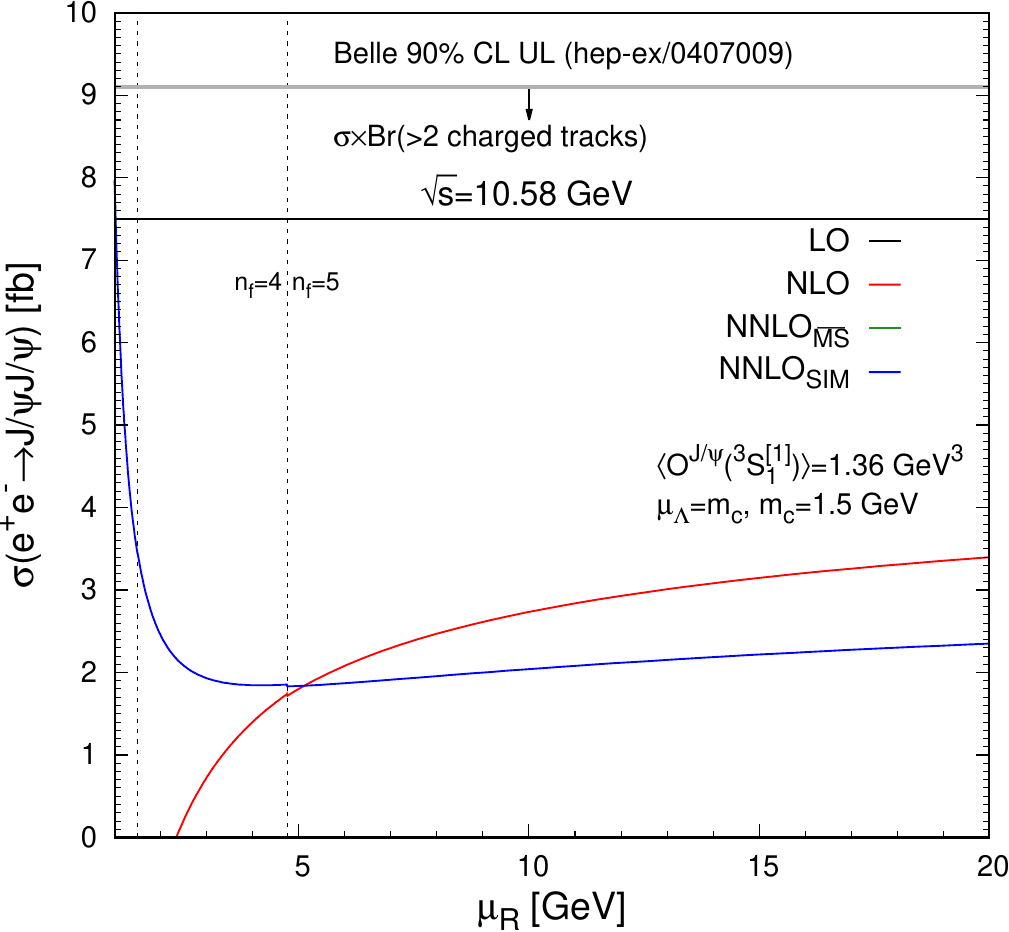}\label{fig:doublecharmonium-b}}
\caption{Renormalization-scale dependence of the total cross sections $\sigma(e^+e^-\to J/\psi\eta_c)$ (left) and $\sigma(e^+e^-\to J/\psi J/\psi)$ (right) at $\sqrt{s}=10.58$ GeV. For $e^+e^-\to J/\psi\eta_c$, the measurements by \texttt{BaBar}~\cite{BaBar:2005nic} (gray band) and \texttt{Belle}~\cite{Belle:2004abn} (purple hatched) correspond to $\sigma(e^+e^-\to J/\psi\eta_c)$ multiplied by the branching fraction of $\eta_c$ into final states with more than two charged tracks. For $e^+e^-\to J/\psi J/\psi$, a $90\%$ CL upper limit on $\sigma(e^+e^-\to J/\psi J/\psi)$ multiplied by the branching fraction of the recoiling $J/\psi$ into final states with more than two charged tracks is set by \texttt{Belle}~\cite{Belle:2004abn}.}
\label{fig:doublecharmonium}
\end{figure}

The NNLO cross section for $e^+e^-\to J/\psi\eta_c$ at $\sqrt{s}=10.58$ GeV has been computed in refs.~\cite{Feng:2019zmt,Huang:2022dfw,Li:2025mng,Chen:2025qgy}. The $\alpha_s$ expansion reads
\begin{equation}
\left.\frac{\sigma_{\mathrm{NNLO}}(e^+e^-\to J/\psi\eta_c)}{\sigma_{\mathrm{LO}}(e^+e^-\to J/\psi\eta_c)}\right|_{\mu_R=\sqrt{s}/2,\sqrt{s}=10.58~\mathrm{GeV}}=1+5.53\alpha_s+\left[17.79_\mathrm{SDC} + 21.46_\mathrm{LDME}\right]\alpha_s^2\,.
\end{equation}
Unlike other processes, no cancellation occurs in the $\alpha_s^2$ coefficient. The logarithmic enhancement in $\log(s/m_c^2)$ has been discussed in refs.~\cite{Jia:2010fw,Bodwin:2014dqa}. The renormalization-scale dependence of the total cross section $\sigma(e^+e^-\to J/\psi \eta_c)$ is shown in fig.~\ref{fig:doublecharmonium-a}. The experimental measurements by \texttt{BaBar}~\cite{BaBar:2005nic} and \texttt{Belle}~\cite{Belle:2004abn} of $\sigma(e^+e^-\to J/\psi\eta_c)\mathrm{Br}(\eta_c\to k~\mathrm{charged~tracks})$ with $k>2$ are compared with theoretical predictions. The NNLO$_{\mathrm{SIM}}$ prediction agrees with the \texttt{BaBar} (\texttt{Belle}) measurement for $\mu_R<4.5$ GeV ($\mu_R<1.5$ GeV), while the NNLO$_{\overline{\mathrm{MS}}}$ prediction consistently undershoots the data.


\subsubsection{$e^+e^-\to J/\psi J/\psi$}

Due to the even charge parity of the final state, the process $e^+e^-\to J/\psi J/\psi$ cannot proceed via a single $s$-channel virtual photon in either QCD or QED. Therefore, its leading production mechanism at $B$-factory energies is through two virtual photons. The NNLO QCD corrections to its cross section have been computed in refs.~\cite{Sang:2023liy,Huang:2023pmn}. The $\alpha_s$ expansion of the total cross section at $\sqrt{s}=10.58$ GeV reads
\begin{equation}
\left.\frac{\sigma_{\mathrm{NNLO}}(e^+e^-\to J/\psi J/\psi)}{\sigma_{\mathrm{LO}}(e^+e^-\to J/\psi J/\psi)}\right|_{\mu_R=\sqrt{s}/2,\sqrt{s}=10.58~\mathrm{GeV}}=1-3.57\alpha_s+\left[
-19.35_{\mathrm{SDC}}
+18.69_{\mathrm{LDME}}\right]\alpha_s^2\,.
\end{equation}
The $\mu_R$ dependence of the cross section $\sigma(e^+e^-\to J/\psi J/\psi)$ is shown in fig.~\ref{fig:doublecharmonium-b}, where the NNLO$_{\overline{\mathrm{MS}}}$ result is negative over the entire range of $\mu_R$ considered. For this process, only a $90\%$ CL upper limit on $\sigma(e^+e^-\to J/\psi J/\psi)\mathrm{Br}(\mathrm{recoiling}~J/\psi \to k~\mathrm{charged~tracks})$ with $k>2$ has been set by \texttt{Belle}~\cite{Belle:2004abn}.

Since the NNLO$_{\overline{\mathrm{MS}}}$ cross section is negative, ref.~\cite{Sang:2023liy} suggests using the electromagnetic decay constant of the $J/\psi$, $f_{J/\psi}$, determined from the experimental value of $\Gamma(J/\psi\to\ell^+\ell^-)$, instead of relying on perturbative calculations for the dominant double-fragmentation contribution. On the other hand, ref.~\cite{Huang:2023pmn} advocates squaring the full amplitude rather than expanding in $\alpha_s$ and truncating at $\mathcal{O}(\alpha_s^2)$ at the squared-amplitude level. This strategy is similar to that used in the $J/\psi\to \gamma\gamma\gamma$ case in ref.~\cite{Zeng:2026ois}. However, squaring the full NNLO$_{\overline{\mathrm{MS}}}$ amplitude, as done in ref.~\cite{Zeng:2026ois}, leads to very large renormalization-scale variations.

\subsubsection{$Z\to \eta_Q\gamma$ and $Z\to V_Q\gamma$}

\begin{figure}[hbt!]
\subfloat[]{\includegraphics[width=0.49\columnwidth,draft=false]{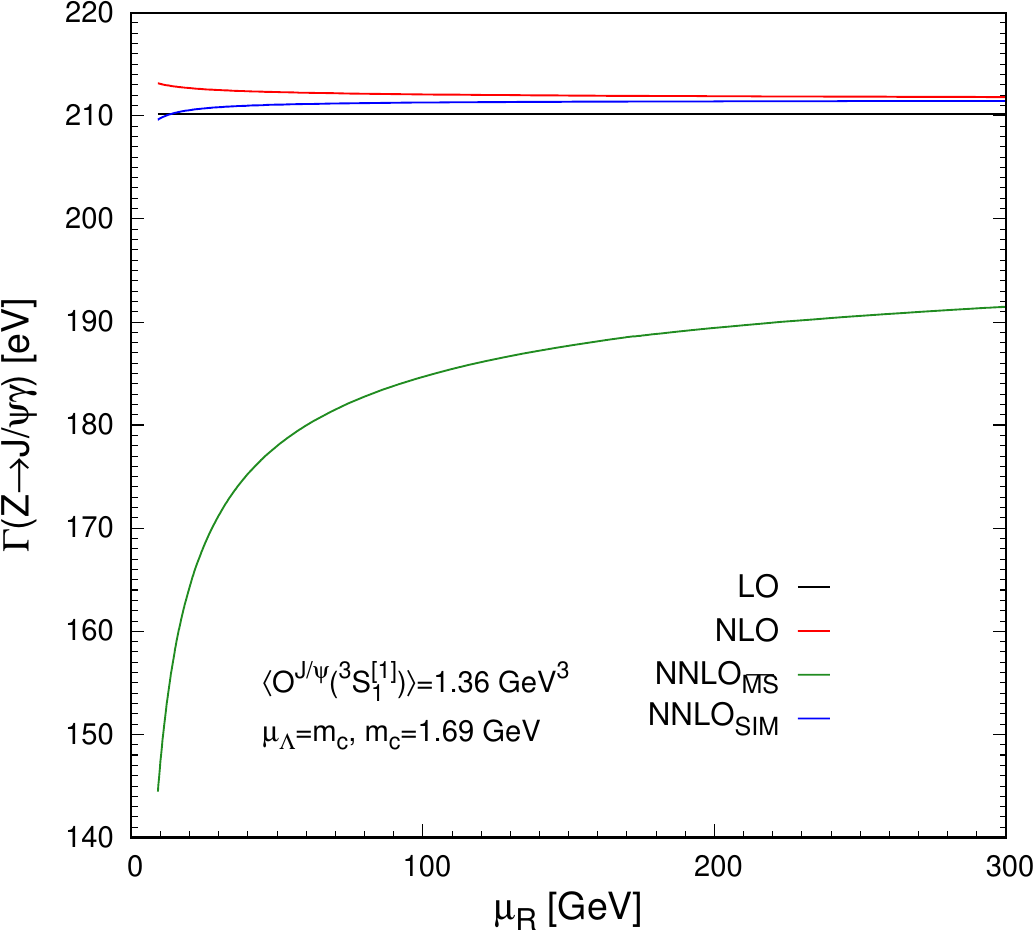}\label{fig:HZdecay-a}}\hfill
\subfloat[]{\includegraphics[width=0.49\columnwidth,draft=false]{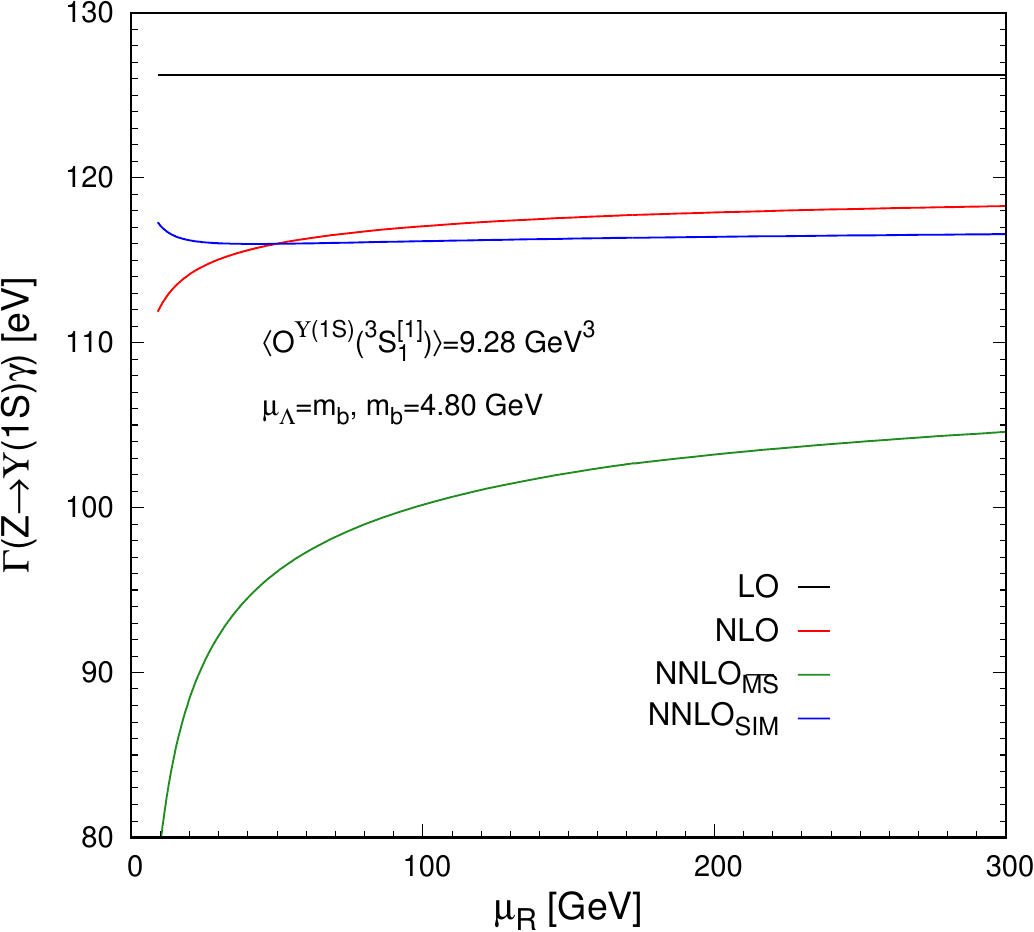}\label{fig:HZdecay-b}}\\
\subfloat[]{\includegraphics[width=0.49\columnwidth,draft=false]{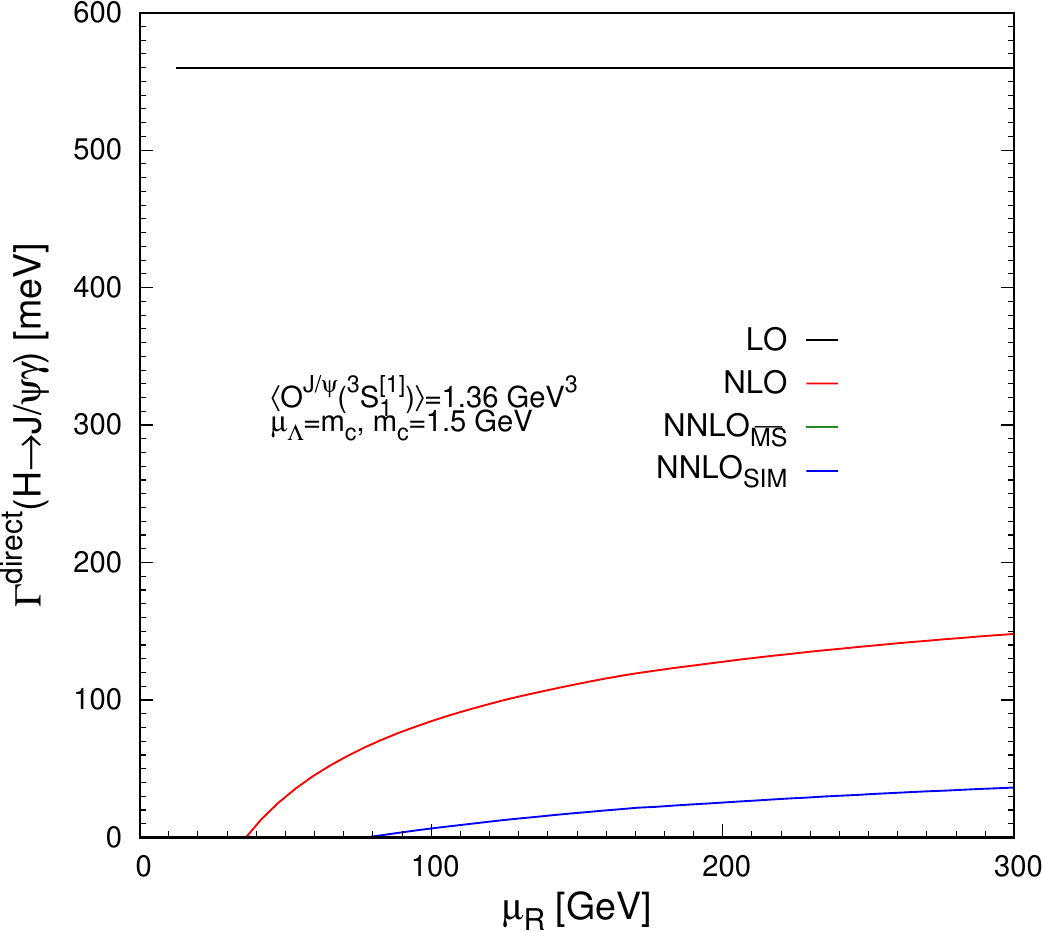}\label{fig:HZdecay-c}}\hfill
\subfloat[]{\includegraphics[width=0.49\columnwidth,draft=false]{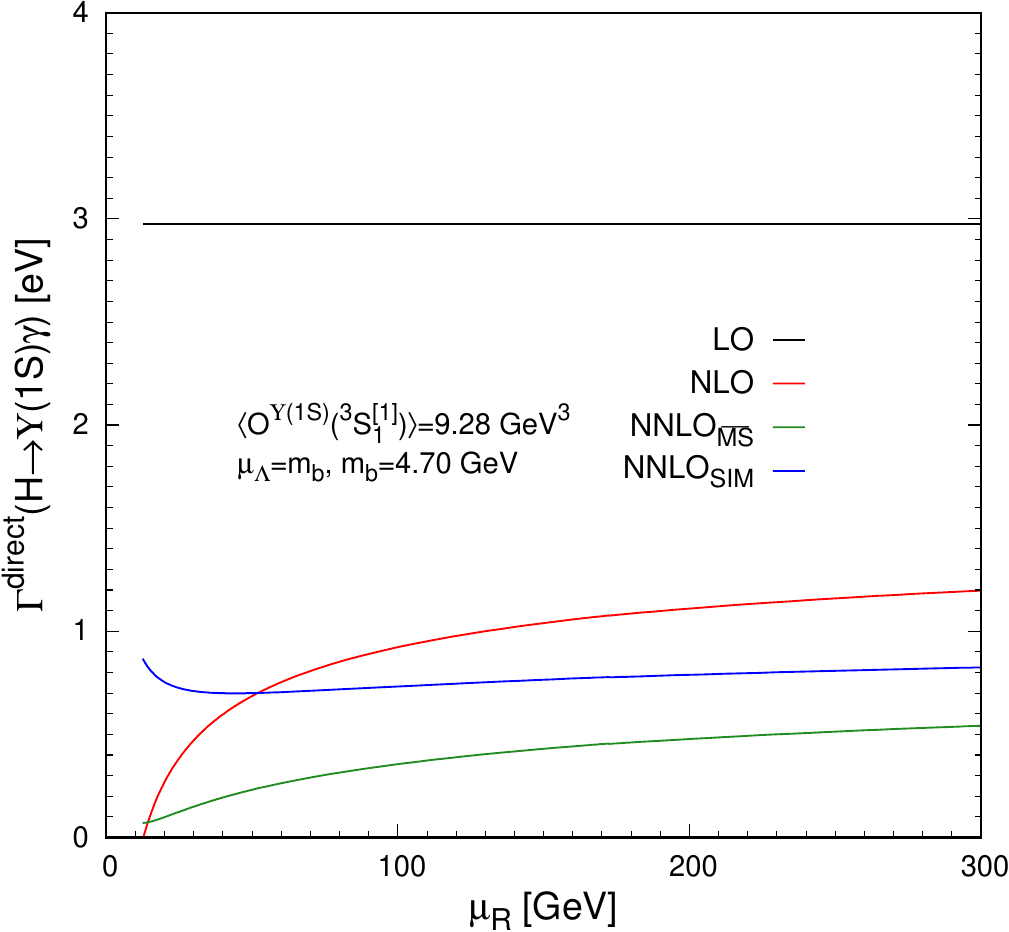}\label{fig:HZdecay-d}}
\caption{Renormalization-scale dependence of the partial decay widths $\Gamma(Z\to J/\psi\gamma)$ (upper left), $\Gamma(Z\to \Upsilon(1S)\gamma)$ (upper right), $\Gamma^{\mathrm{direct}}(H\to J/\psi\gamma)$ (lower left), and $\Gamma^{\mathrm{direct}}(H\to \Upsilon(1S)\gamma)$ (lower right). For Higgs decays, only the Yukawa-coupling-dependent direct contributions are included.}
\label{fig:HZdecay}
\end{figure}

NNLO QCD corrections to the $Z$ boson decay into quarkonium plus a photon have been computed in ref.~\cite{Sang:2023hjl}, where the charm- and bottom-quark masses are fixed to $m_c=1.69$ GeV and $m_b=4.80$ GeV. This setup differs slightly from eq.~\eqref{eq:inputparameters}. We adopt these mass values to assess the perturbative convergence. For spin-triplet quarkonium, the $\alpha_s$ expansion reads
\begin{equation}
\left.\frac{\Gamma_{\mathrm{NNLO}}(Z\to V_Q \gamma)}{\Gamma_{\mathrm{LO}}(Z\to V_Q \gamma)}\right|_{\mu_R=m_Z/2}=1+\left(\begin{array}{c}+0.08_{n_f=3}\\
-0.62_{n_f=4}
\end{array}\right)\alpha_s+\left[\left(\begin{array}{c}-9.65_{\mathrm{SDC},n_f=3}\\
-9.43_{\mathrm{SDC},n_f=4}
\end{array}\right)+9.35_{\mathrm{LDME}}\right]\alpha_s^2\,,
\end{equation}
while for spin-singlet states it reads
\begin{equation}
\left.\frac{\Gamma_{\mathrm{NNLO}}(Z\to\eta_Q \gamma)}{\Gamma_{\mathrm{LO}}(Z\to\eta_Q \gamma)}\right|_{\mu_R=m_Z/2}=1+\left(\begin{array}{c}+0.66_{n_f=3}\\
-0.08_{n_f=4}
\end{array}\right)\alpha_s+\left[\left(\begin{array}{c}-12.95_{\mathrm{SDC},n_f=3}\\
-13.08_{\mathrm{SDC},n_f=4}
\end{array}\right)+12.11_{\mathrm{LDME}}\right]\alpha_s^2\,.
\end{equation}
The $\alpha_s$ coefficients are relatively small due to accidental cancellations between the non-logarithmic terms and the contribution $\frac{\alpha_s}{\pi}C_F\left(\frac{3}{2}-\log 2\right)\log{\left(\frac{m_Z^2}{4m_Q^2}\right)}$ in the leading-power expansion in $4m_Q^2/m_Z^2$. The Sudakov logarithm $\log{\left(\frac{m_Z^2}{4m_Q^2}\right)}$ has been resummed to next-to-leading logarithmic accuracy in ref.~\cite{Sang:2023hjl} using the light-cone factorization approach. The renormalization-scale dependence of $\Gamma(Z\to J/\psi\gamma)$ and $\Gamma(Z\to \Upsilon(1S)\gamma)$ is shown in figs.~\ref{fig:HZdecay-a} and \ref{fig:HZdecay-b}. Using $\mu_R=\xi_R m_Z/2$ with $\xi_R\in\left\{1,1/2,2\right\}$, the partial widths are also presented in tab.~\ref{tab:widthZ2VQetaQa}, together with the strongest $95\%$ CL upper limits set by the ATLAS~\cite{ATLAS:2022rej} and CMS~\cite{CMS:2024hhg} collaborations.

\begin{table}[htpb!]
\centering
\tabcolsep=3.5mm
\renewcommand*{\arraystretch}{1.3}
\vspace{0.2cm}
\begin{tabular}{|c|c|c|c|c|c|} \hline
 \multirow{2}{*}{Observable} & \multirow{2}{*}{Experiment} & \multicolumn{4}{c|}{Theory} \\\cline{3-6}
 & & LO & NLO & NNLO$_{\overline{\mathrm{MS}}}$ & NNLO$_{\mathrm{SIM}}$ \\ \hline
$\Gamma(Z \to J/\psi \gamma)$ [eV] & $< 1497$~\cite{CMS:2024hhg}  &  $210.2$ & $212.3^{+0.3}_{-0.2}$ & $176.9_{-10.2}^{+7.0}$ & $211.0^{+0.2}_{-0.4}$ \\
$\Gamma(Z \to \Upsilon(1S) \gamma)$ [eV] & $< 2745$~\cite{ATLAS:2022rej} &  $126.2$ & $115.8^{+1.1}_{-1.4}$ & $95.5^{+4.2}_{-5.7}$ & $116.0^{+0.1}_{-0.0}$ \\
$\Gamma(Z \to \eta_c \gamma)$ [eV] &   &  $24.1$& $26.2^{+0.3}_{-0.2}$ & $20.6^{+1.1}_{-1.6}$ & $25.7^{+0.1}_{-0.2}$\\
$\Gamma(Z \to \eta_b \gamma)$ [eV] &   &  $61.6$ & $60.9^{+0.1}_{-0.1}$ & $46.9^{+2.8}_{-4.0}$ & $59.9^{+0.2}_{-0.3}$ \\\hline
\end{tabular}
\caption{Comparison of theoretical predictions for quarkonium plus a photon production in $Z$ boson decays with experimental measurements~\cite{CMS:2024hhg,ATLAS:2022rej}. The charm- and bottom-quark masses are taken as $m_c=1.69$ GeV and $m_b=4.80$ GeV. The central renormalization scale is chosen as $\mu_{R,0}=m_Z/2$, and the quoted theoretical uncertainties correspond to scale variations obtained by varying $\mu_R$ by a factor of two around $\mu_{R,0}$.\label{tab:widthZ2VQetaQa}
}
\end{table}

\subsubsection{$H\to V_Q\gamma$}

The exclusive rare decay modes of the Higgs boson into quarkonium plus a photon are interesting probes of the Yukawa couplings. Therefore, both the ATLAS and CMS collaborations continue to measure these channels at the LHC. The current $95\%$ CL upper limits on the branching ratios $\mathrm{Br}(H\to \psi(nS)\gamma)$ ($n=1,2$) and $\mathrm{Br}(H\to \Upsilon(nS)\gamma)$ ($n=1,2,3$) are reported in refs.~\cite{ATLAS:2015vss,ATLAS:2018xfc,ATLAS:2022rej,CMS:2018gcm,CMS:2024hhg}. The amplitudes of these decay processes are usually decomposed into a direct contribution, proportional to the constituent heavy-quark Yukawa coupling, and an indirect contribution arising from $H\to \gamma^*\gamma\to V_Q\gamma$. NNLO QCD corrections are known only for the direct amplitudes~\cite{Jia:2024ini}. We therefore focus exclusively on the direct contribution, neglecting both the indirect contribution and the interference. The results of ref.~\cite{Jia:2024ini} are presented with fixed masses $m_c=1.5$ GeV for $V_Q=J/\psi$ and $m_b=4.7$ GeV for $V_Q=\Upsilon(1S)$. We adopt these mass values to assess the perturbative convergence of the direct partial widths $\Gamma^{\mathrm{direct}}(H\to J/\psi\gamma)$ and $\Gamma^{\mathrm{direct}}(H\to \Upsilon(1S)\gamma)$. The $\alpha_s$ series reads
\begin{equation}
\left.\frac{\Gamma_{\mathrm{NNLO}}^\mathrm{direct}(H\to V_Q \gamma)}{\Gamma_{\mathrm{LO}}^\mathrm{direct}(H\to V_Q \gamma)}\right|_{\mu_R=m_H/2}=1+\left(\begin{array}{c}-7.29_{n_f=3}\\
-5.93_{n_f=4}
\end{array}\right)\alpha_s+\left[\left(\begin{array}{c}-20.37_{\mathrm{SDC},n_f=3}\\
-12.31_{\mathrm{SDC},n_f=4}
\end{array}\right)+9.35_{\mathrm{LDME}}\right]\alpha_s^2\,.
\end{equation}
Large negative NLO QCD corrections stem from the Sudakov logarithm $-\frac{\alpha_s}{\pi}C_F\log(2)\log{\left(\frac{m_H^2}{4m_Q^2}\right)}$. Such logarithms are smaller in the bottomonium case than in the charmonium case. Their further improvements will require Sudakov logarithm resummation. The renormalization-scale dependence of the two partial decay widths is shown in figs.~\ref{fig:HZdecay-c} and \ref{fig:HZdecay-d}. In the case of $H\to J/\psi\gamma$, the NNLO$_{\overline{\mathrm{MS}}}$ width is negative and is therefore not visible.

\subsection{Theory-Data Comparison}

\begin{figure}[hbt!]

\includegraphics[width=0.95\columnwidth,draft=false]{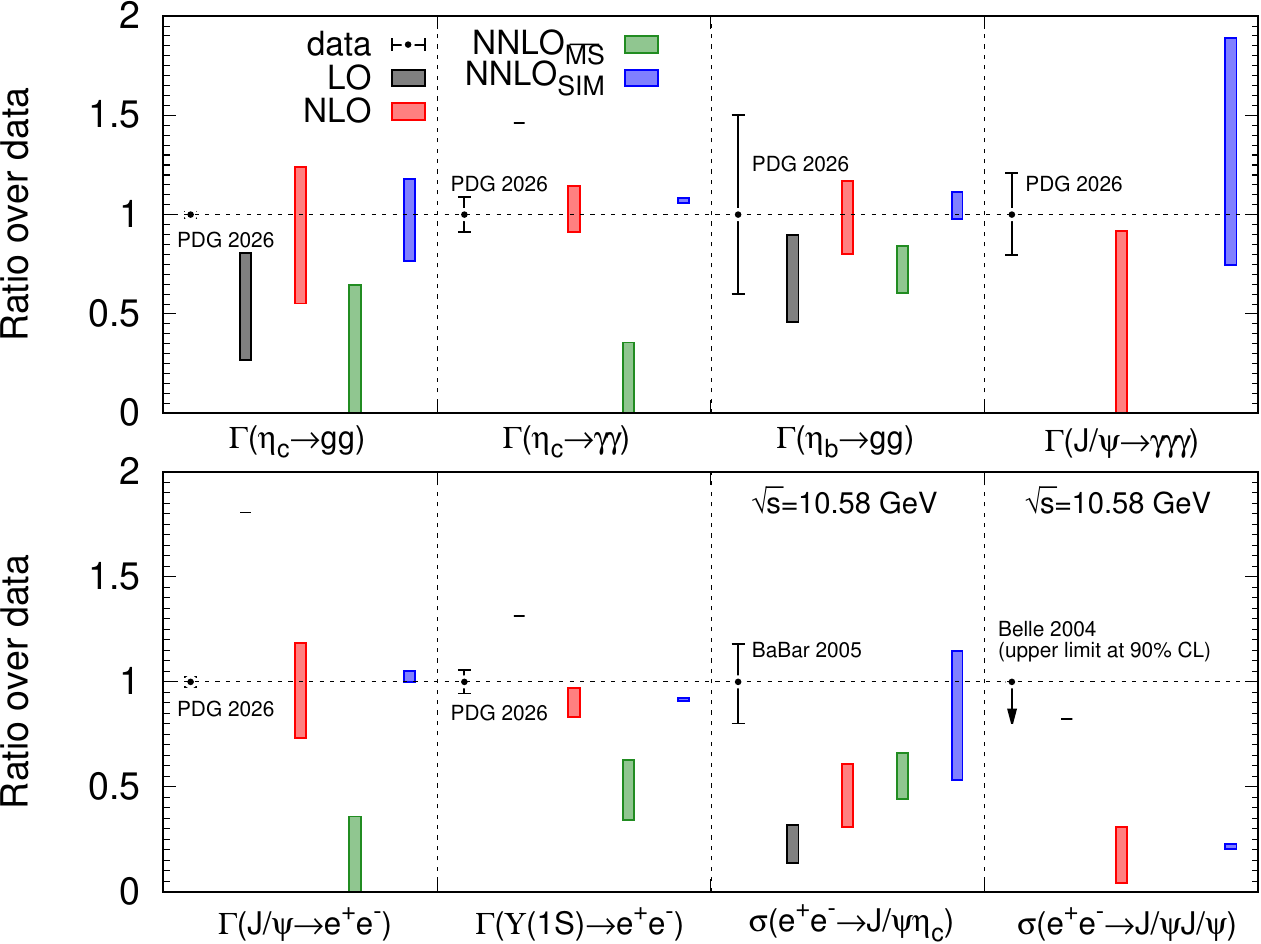}
\caption{Ratios of theoretical predictions to the central values of the experimental data. The \texttt{PDG}, \texttt{BaBar}, and \texttt{Belle} data are taken from refs.~\cite{ParticleDataGroup:2026aaa}, \cite{BaBar:2005nic}, and \cite{Belle:2004abn}, respectively.}
\label{figwidthxs} 
\end{figure}

To facilitate the comparison between theory and data, we present the results of tab.~\ref{tab:widthxs} in fig.~\ref{figwidthxs}, where the theoretical predictions are normalized to the central values of the experimental data~\cite{ParticleDataGroup:2026aaa,BaBar:2005nic,Belle:2004abn}.

\end{document}